\documentclass[journal]{IEEEtran}

\usepackage{indentfirst}
\usepackage{graphicx}
\usepackage{amsmath}
\usepackage{amssymb}
\usepackage{amsfonts}
\usepackage{mathrsfs}
\usepackage{leftidx}
\usepackage{color}
\usepackage{amsmath}
\usepackage{arydshln}
\usepackage{amsthm}
\usepackage{ragged2e}
\usepackage{cite}
\usepackage{enumerate}
\usepackage{longtable}
\usepackage{float}
\usepackage{stfloats}
\usepackage{hyperref }
\usepackage{algpseudocode}
\usepackage{algorithm}
\usepackage{subfigure}
\usepackage{tabularx}

\theoremstyle{plain}

\theoremstyle{definition}

\usepackage{glossaries}

\usepackage[table,usenames,dvipsnames]{xcolor}
\definecolor{aa}{RGB}{175,238,238}
\definecolor{bb}{RGB}{255,255,255}
\usepackage{bm}
\usepackage{makecell}

\usepackage{siunitx}

\begin{document}

\title{Importance-Aware Robust Semantic Transmission for LEO Satellite-Ground Communication}

\author{Hui Cao, Rui Meng,~\IEEEmembership{Member,~IEEE,} Xiaodong Xu,~\IEEEmembership{Senior Member,~IEEE,} 

Shujun Han,~\IEEEmembership{Member,~IEEE,} 
and Ping Zhang,~\IEEEmembership{Fellow,~IEEE}

\thanks{This work was supported in part by the National Key Research and Development Program of China under Grant 2020YFB1806905; in part by the National Natural Science Foundation of China under Grant 62501066 and under Grant U24B20131; and in part by the Beijing Natural Science Foundation under Grant L242012. \textit{(Corresponding author: Rui Meng and Xiaodong Xu.)}}
\thanks{Hui Cao, Rui Meng, Xiaodong Xu, Shujun Han, and Ping Zhang are with the State Key Laboratory of Networking and Switching Technology, Beijing University of Posts and Telecommunications, Beijing 100876, China (e-mail: caohui@bupt.edu.cn; buptmengrui@bupt.edu.cn;  xuxiaodong@bupt.edu.cn; hanshujun@bupt.edu.cn; pzhang@bupt.edu.cn).}

}



\maketitle

\begin{abstract}
Satellite-ground semantic communication is anticipated to serve a critical role in the forthcoming sixth-generation (6G) mobile networks. Nonetheless, task-oriented data transmission in such systems remains a formidable challenge, primarily due to the dynamic nature of Signal-to-Noise Ratio (SNR) fluctuations and the stringent bandwidth limitations inherent to Low Earth Orbit (LEO) satellite channels. In response to these constraints, we propose an Importance-Aware Robust Semantic Transmission (IRST) framework, specifically designed for scenarios characterized by bandwidth scarcity and channel variability. The IRST scheme begins by applying a segmentation model enhancement algorithm to improve the granularity and accuracy of semantic segmentation. Subsequently, a task-driven semantic selection method is employed to prioritize the transmission of semantically vital content based on real-time Channel State Information (CSI). Furthermore, the framework incorporates a stack-based, SNR-aware channel codec capable of executing adaptive channel coding in alignment with SNR variations. Comparative evaluations across diverse operating conditions demonstrate the superior performance and resilience of the IRST model relative to existing benchmarks. The code is
available at https://github.com/lightwindy-ch/IRST.git.
\end{abstract}


\section{Introduction}
% \subsection{Background}
\IEEEPARstart{A}{s} 
the global demand for high-speed, low-latency, wide-coverage communications continues to grow, satellite-ground communication will play an indispensable role in sixth-generation (6G) mobile networks \cite{cao2025exploring,fan2025generative}. Compared to terrestrial networks, satellite-ground communication offers broader coverage, particularly in regions with limited infrastructure such as mountains, oceans, and polar areas, thereby enhancing communication accessibility \cite{lin2024satellite-mec}. Moreover, satellite-ground communication, particularly Low Earth Orbit (LEO) satellite communication, offers several compelling advantages, including reduced communication latency, lower path loss, enhanced system capacity, and greater operational flexibility \cite{al2023satellite, wang2025multiple,liu2024history}. Therefore, satellite-ground communication plays an important role in emerging application scenarios, including the global Internet of Things (IoT), disaster response, and telemedicine \cite{wang2020realzing}.

Despite its potential, satellite-ground communication faces several critical challenges.
Firstly, due to the high-velocity motion of the satellites, the channel state of the satellite-ground link undergoes rapid variation. These fluctuations in channel conditions result in significant degradation of signal quality, negatively impacting the performance of modulation and demodulation processes. Consequently, the system experiences an elevated bit error rate, which compromises overall communication reliability \cite{chen2024traffic}.
Secondly, satellite-ground communication systems face inherent limitations in bandwidth resources. In such constrained environments, insufficient onboard computing power can hinder the efficient allocation and utilization of available bandwidth. This inefficiency may result in the waste or congestion of spectrum resources, ultimately compromising the Quality of Service (QoS) guarantees required for reliable transmission \cite{araniti2015effective}.

% \subsection{Satellite-Ground Semantic Communication (SG-SemCom)}

To address these challenges, Semantic Communication (SemCom) has become a promising solution \cite{rong2025semantic,meng2025survey}. Unlike traditional communication systems that primarily aim to transmit raw bits accurately, SemCom emphasizes the transmission of meaning and the fulfillment of communicative intent. In addition, instead of ensuring only complete data reception, SemCom prioritizes the receiver’s ability to correctly interpret and reconstruct the intended semantic content of the sender \cite{shannon1948mathematical}. 
Currently, SemCom has been widely studied for ground networks. For example, Wang et al. \cite{wang2025latent} propose a SemCom framework based on latent diffusion model, combined with Variational Autoencoder (VAE) for semantic compression, to achieve robust transmission without additional training. Additionally, Tong et al.  \cite{tong2025alternate} propose  an alternating learning mechanism to improve the quality of image semantic restoration by combining compressed sensing and lightweight U-Net denoising structure. 

% In addition, methods such as generative adversarial networks (GAN) \cite{chen2018image}, Transformers \cite{yang2024swinjscc}, and joint source channel coding (JSCC) \cite{bourtsoulatze2019deep} are also commonly employed for denoising in SemCom.

SemCom has significant potential in improving the efficiency and reliability of satellite-ground communication systems.
Firstly, in satellite links characterized by unstable channel conditions and transmission errors, SemCom can facilitate information recovery through semantic reasoning, thus improving transmission robustness \cite{chen2024semantic}. Secondly, by transmitting only essential and meaningful information, SemCom significantly reduces data load across satellite links, contributing to efficient utilization of limited bandwidth resources, especially critical in LEO satellite systems \cite{deng2024semantic}. Lastly, by focusing on the transmission of task intent rather than raw bit streams, SemCom aligns well with low-latency, high-efficiency edge computing and intelligent decision-making scenarios, ultimately advancing the cognitive and adaptive capabilities of satellite-ground communication systems \cite{lu2025important}.

Driven by these advantages, researchers have employed SemCom in satellite-ground systems. As mentioned above, the dynamic change of the satellite-ground channel affects the stability of communication and presents the challenge of deterioration of link reliability. To address the issue, Chen et al. \cite{chen2024semantic} propose a method combining SemCom with Orthogonal Time-Frequency Space (OTFS) modulation to effectively alleviate the severe Doppler effect and improve transmission efficiency. Furthermore, Jiang et al. \cite{jiang2025feature} use multirandom noise and multichannel mulation to improve the robustness of the transmission data during the training process, so that a single training model can be adapted to remote sensing applications with different Signal-to-Noise Ratio (SNR) and harsh environments.Yuan et al. \cite{yuan2024deep} propose a ModNet channel guided by Channel State Information (CSI) that aligns the semantics of the image with the transmission conditions, optimizing the trade-off between the quality of the reconstruction and the utility of the network. Moreover, Jiang et al. \cite{jiang2025semantic} investigate denoising and restoration strategies for satellite-based SemCom networks. Among these, Foundation Model (FM)-based segmentation and reconstruction techniques demonstrate robust performance in preserving semantic features under high-noise conditions. In addition, advances have been made in optimizing the selection of gateway hop relays for low-SNR ground users, coupled with the integration of gateway-based denoising mechanisms \cite{nguyen2025semantic}. 

Furthermore, the bandwidth resource of the satellite-ground communication system is limited, reducing the utilization efficiency. To address this issue, semantic importance is a feasible method. Sun et al. \cite{sun2022deep} calculate semantic importance by analyzing the gradient of task perception results, and incorporate semantic weighting into the loss function design. To overcome related limitations, various semantic awareness strategies have emerged, emphasizing data utility and relevance across interconnected devices \cite{lu2025hyper}. Approaches include the Feature Allocation for Semantic Transmission (FAST) framework for feature evaluation \cite{zhou2025feature}, entropy-based models for latent semantic discrimination \cite{xiao2023wireless}, and gradient-based methods for semantic concept importance \cite{liu2024adaptable, wang2024feature}. Features exceeding a predefined weight threshold are preserved, while others are discarded.

However, the aforementioned methods are not directly applicable to the satellite-ground scenario. First, semantic evaluation models trained in terrestrial environments often struggle to generalize to the unique noise and interference characteristics of satellite-ground channels, leading to biased semantic weight assessments and degraded transmission efficiency \cite{jiang2025semantic}. Second, many semantic awareness strategies (such as gradient-based analysis and entropy modeling) are computationally intensive \cite{xu2024semantic}, while satellites typically operate under constrained processing capabilities and energy budgets, limiting their ability to perform real-time reasoning or update model parameters \cite{chou2024cognitive}.

Therefore, although SemCom has made significant progress in terrestrial networks, its application in satellite-ground communication scenarios still faces many challenges. Especially in terms of dynamic noise reduction mechanism, satellite-ground channel adaptability, and multi-dataset generalization ability, the existing methods have obvious limitations. Most existing methods rely on fixed semantic compression and recovery structures, lacking adaptability to real-time channel variations. Secondly, in the satellite-ground link, channel noise may lead to misjudgment of semantic features, thereby incorrectly discarding key semantic information and affecting task completion. Meanwhile, inconsistent definitions of semantic importance across datasets hinder the development of a unified selection mechanism. Finally, most methods rely on offline training and lack a lightweight online update mechanism, which cannot adapt to task changes or environmental disturbances.

\begin{table*}[]
\caption{Comparisons between the proposed IRST scheme with existing schemes}
\centering
\begin{tabular}{ccccc}
\hline
Ref. & Semantic Importance &  Dynamic Denoising & Suitable for Satellite-Ground Scenario  & Validation Using Multiple Datasets \\ \hline
\cite{chen2024semantic}   & ×               & \checkmark                      & \checkmark             & ×  \\
\cite{jiang2025feature}   & \checkmark                   & \checkmark                                      & ×            & \checkmark                        \\
\cite{yuan2024deep}     & ×                   & \checkmark                                              & \checkmark      & ×        \\
\cite{jiang2025semantic}     & \checkmark                  & ×                                          & \checkmark       & \checkmark             \\
\cite{nguyen2025semantic}     & ×                  & \checkmark                                     & ×       & \checkmark                       \\
\cite{sun2022deep}     & \checkmark                  & \checkmark                               & ×              & ×                     \\
\cite{xu2024semantic}     & \checkmark                  & ×                                          & ×        & ×                 \\
\cite{yang2023computation}     & \checkmark                  & ×                                  & \checkmark            & ×              \\
\cite{xu2018toward}     & \checkmark                  & ×                                       & ×          & ×            \\
\cite{furutanpey2025fool}     & \checkmark                  & ×                                       & ×         & \checkmark                  \\
\cite{bourtsoulatze2019deep}      & ×
& \checkmark                                      & ×        & ×                  \\
\cite{yang2024swinjscc}      & ×
&\checkmark                                      &×        & \checkmark                  \\
\cite{okhotin2023star}    & ×
& \checkmark                                      & ×         &  ×                \\
Proposed IRST     & \checkmark                  & \checkmark                                        & \checkmark       & \checkmark                \\
                         \hline
\end{tabular}
\label{Table.1}
\end{table*}

To address the aforementioned challenges, we propose an Importance-Aware Robust Semantic Transmission (IRST) model that enables content-aware transmission under dynamic SNR conditions. The framework integrates three modules: semantic segmentation, semantic selection, and data transmission, achieving end-to-end optimization from semantic extraction to channel adaptation. This framework not only improves semantic recovery capabilities in low SNR environments, but also significantly reduces redundant data transmission, and has stronger task adaptability and cross-scenario generalization capabilities. Meanwhile, in contrast to existing semantic perception methods that are computationally intensive, the proposed IRST framework achieves substantial complexity reduction through a modular architecture and dynamic adaptation strategies. Specifically, the semantic selection module employs a lightweight evaluation function to identify task-relevant regions, thereby circumventing the need for gradient-based optimization or entropy-based modeling. Furthermore, the SNR-adaptive channel codec incorporates a hierarchical activation mechanism that selectively engages deeper network layers only under low-SNR conditions, effectively conserving computational resources in high-SNR scenarios. Finally, the framework supports end-to-end inference without requiring frequent model updates or retraining, rendering it well-suited for deployment in resource-constrained satellite platforms.

Table \ref{Table.1} compares the proposed scheme with existing representative SemCom schemes.
The main contributions of this paper are as follows.

\begin{itemize}
\item We develop an IRST scheme for satellite-ground communication scenarios with limited bandwidth and dynamic channel variability. This scheme decomposes the image through semantic segmentation and prioritizes task-relevant image content for transmission. It actively adapts to real-time SNR fluctuations to enhance transmission efficiency and preserve semantic fidelity under variable channel conditions.
\item We propose a Segmentation Model Enhancement (SME) method to address the issue of poor segmentation performance in edge regions commonly observed in current semantic segmentation models. This method aims to classify and correct isolated pixels, thereby improving boundary delineation and overall segmentation accuracy.
\item We design a task-oriented semantic selection algorithm that actively filters and prioritizes the transmitted content to meet diverse task requirements and adapt to varying channel conditions. 
% By dynamically aligning transmission strategies with semantic importance and channel capacity, our approach ensures the efficient delivery of mission-relevant information. 
We further develop a specialized SNR-adaptive codec to accommodate varying SNR environments. 
% Using a stacked architectural design, it flexibly encodes and decodes across multiple channel layers. 
It dynamically selects and processes appropriate layers based on real-time SNR conditions, thereby optimizing transmission efficiency and enhancing reconstruction performance. 
\item The simulation results on open source datasets \cite{shao2018performance,cheng2014multi} demonstrate the superior performance of the proposed IRST scheme compared to existing approaches, including Deep JSCC \cite{bourtsoulatze2019deep}, WITT \cite{yang2024swinjscc}, and SS-DDPM \cite{okhotin2023star}. Furthermore, ablation experiments validate the effectiveness of the proposed semantic segmentation, semantic selection, and data transmission modules.
\end{itemize}

The rest of this paper is organized as follows. Section \ref{section2} provides the system model and overviews the proposed IRST scheme. Sections \ref{section3}, \ref{section4}, and \ref{section5} present the proposed semantic segmentation, task-oriented semantic selection, and dynamic SNR-based transmission modules, respectively. Section \ref{section6} gives the simulation results and analysis. Section \ref{section7} concludes this paper.

\section{System Model}
\label{section2}

\subsection{Network Model}
\begin{figure}[!h]
\centering
\includegraphics[width=3in]{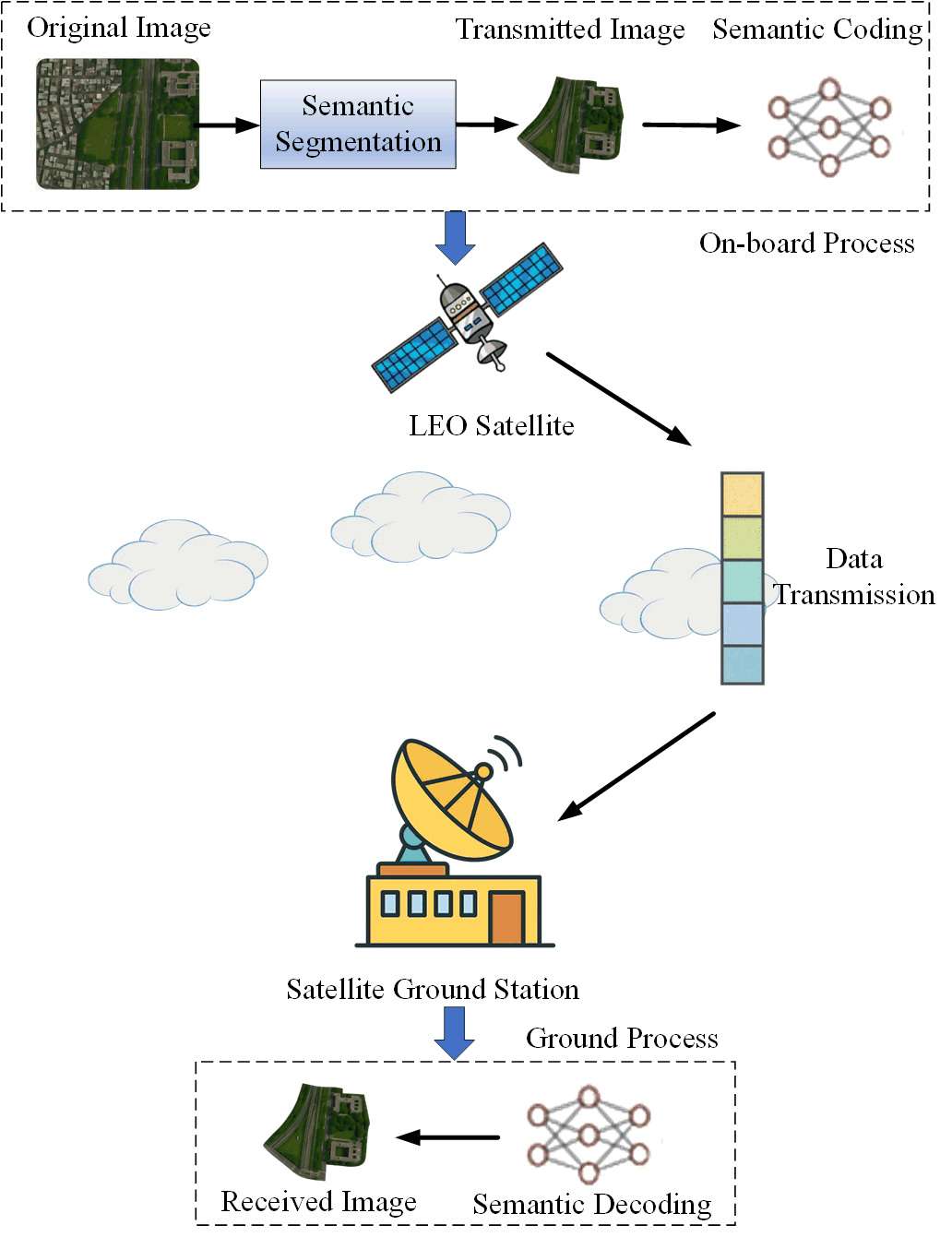}
\caption{The system model of satellite-ground transmission, including LEO satellite, transmission link, and satellite ground station, for satellite-ground image transmission.}
\label{fig1}
\end{figure}

We consider a scenario where images are transmitted from satellite to ground, as shown in Figure \ref{fig1}. The components involved are as follows.

\begin{itemize}
\item LEO satellite: Satellites rotate around the Earth to cover different regions. The satellite observes the image generated by the ground data and selects the content in the image according to different mission requirements. The satellite encodes the content that needs to be transmitted, including semantic coding and channel coding, to improve transmission efficiency and data quality.
\item Transmission link: The link between the satellite and the ground equipment through the wireless channel will be affected by environmental factors such as clouds and electromagnetic interference, thereby reducing the quality of the signal. In addition, due to the movement of LEO satellites, the transmission path and conditions will change with the change of satellite position, so the SNR of the transmission link is in dynamic change.
\item Satellite ground station: The satellite ground station receives the signal sent by the satellite through the antenna and performs channel decoding and semantic decoding in turn to recover the transmitted image.
\end{itemize}

\subsection{Channel Model}
Unlike the traditional terrestrial network, in the satellite-ground transmission scenario, the channel between the satellite and the ground user is not only affected by the distance loss caused by the Line-of-Sight (LoS) propagation, but also by atmospheric fading, the shadow effect caused by obstacles, and the scattering. We use the Shadowed-Rician (SR) fading channel, which is widely used in the S-band and ka-band satellite channels for data transmission \cite{adbi2003new,hong2024age,yuan2024deep}. In this paper, the satellite uses a single transmitting antenna. According to \cite{adbi2003new}, the Probability Density Function (PDF) of the SR channel gain ${f_h}(r)$ is expressed as

\begin{equation}
\label{1}
\begin{array}{l}
{f_h}(r) = {\left( {\frac{{2{b_0}m}}{{2{b_0}m + \Omega }}} \right)^m}\frac{1}{{2{b_0}}}\exp \left( { - \frac{r}{{2{b_0}}}} \right) \\ \cdot 
{}_1{F_1}\left( {m,1,\frac{{\Omega r}}{{2{b_0}(2{b_0}m + \Omega )}}} \right),
\end{array}
\end{equation}

\noindent where ${b_0}$, $m$, and $\Omega$ denote the average power of the scatter component, the Nakagami-$m$ parameter, and the LoS component, respectively. ${}_1{F_1}(a,b,c)$ is the first kind of confluent hypergeometric function. In particular, for the parameter $m$, when $m$ is small, it indicates that there are more obstacles and shadows between the satellite and the ground, which may not meet the conditions for establishing the LoS link. As the parameter $m$ approaches infinity, it signifies that the satellite-to-ground link experiences minimal obstruction, effectively establishing a LoS communication condition. Under this scenario, the statistical characteristics of the SR distribution converge toward those of the Rayleigh distribution, as the influence of multipath fading becomes predominant in the absence of significant shadowing effects \cite{yuan2024deep}.

\subsection{Overview of the Proposed Importance-Aware Robust Semantic Transmission (IRST) Framework}

\begin{figure*}[!t]
\centerline{\includegraphics[width=1\textwidth]{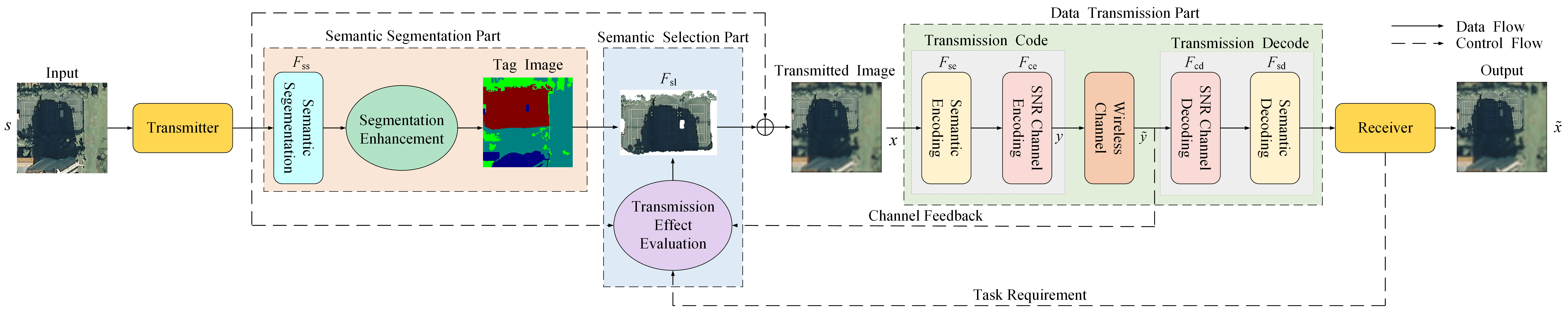}}
\caption{The architecture of the proposed IRST, which is mainly composed of semantic segmentation, semantic selection, and data transmission modules.}
\label{fig2}
\end{figure*}

As presented in Figure \ref{fig2}, the proposed IRST scheme includes the following modules.

\subsubsection{Semantic Segmentation Module}
An image source of Red, Green, Blue (RGB) $\mathbf{s}$ is input to the semantic segmentation model. The model performs semantic segmentation and generates the corresponding output $\mathbf{s_{ss}}$. This process is modeled as
\begin{equation}
\label{2}
\begin{array}{l}
\mathbf{s_{ss}} = {f_{ss}}(\mathbf{s},{\mathbf{\theta _{ss}}}),
\end{array}
\end{equation}
where ${\mathbf{\theta _{ss}}}$ are the model parameters for semantic segmentation. 

\subsubsection{Semantic Selection Module}
Based on task requirements and current SNR conditions, the semantic segmentation image $\mathbf{s_{ss}}$ is utilized. The content of the original image $\mathbf{s}$ is analyzed to identify the most relevant regions. Noncritical regions are then blurred to prioritize the essential image information. The process is modeled as
\begin{equation}
\label{3}
\begin{array}{l}
\mathbf{x} = {f_{sl}}(\mathbf{s},\mathbf{s_{ss}},\gamma,q),
\end{array}
\end{equation}
where $x$ is the image that needs to be transmitted, and $\gamma$ and ${q}$ are the current SNR in wireless channel and the task requirements, respectively.

\subsubsection{Data Transmission  Module}
The semantic encoder and decoder utilize the Swin Transformer model \cite{liu2021swin} as the pre-trained model. It is built on the multi-head attention mechanism. Meanwhile, the channel encoder and decoder employ a segmented architecture, and the segmentation is based on SNR conditions. The expression of this process are 
\begin{equation}
\label{4}
\begin{array}{l}
{\mathbf{x_{se}}} = {f_{se}}(\mathbf{x},\mathbf{{\theta _{se}}})
\end{array}
\end{equation}
and
\begin{equation}
\label{5}
\begin{array}{l}
\mathbf{y} = {f_{ce}}(\mathbf{{x_{se}}},\mathbf{\theta _{ce}}),
\end{array}
\end{equation}
where $\mathbf{x_{se}}$ is the result of the semantic encoder. $\mathbf{\theta _{se}}$ and $\mathbf{\theta _{ce}}$ are the model parameters of the semantic encoder and the channel encoder, respectively. $\mathbf{y}$ is the signal after passing through the channel encoder.
Beyond the encoder, $\mathbf{y}$ will be sent to the satellite-ground channel for transmission. Due to the existence of channel fading and noise, the signal $\mathbf{y}$ will change to $\mathbf{\tilde y}$ after passing through the channel. The fading process is modeled as 

\begin{equation}
\label{6}
\begin{array}{l}
\mathbf{\tilde y} = h\mathbf{y} + n,
\end{array}
\end{equation}
where $\mathbf{n}$ is random Gaussian noise and $\mathbf{n}\sim(0,{\sigma ^2})$, and $\mathbf{h}$ is the gain of the channel. 
Finally, $\mathbf{\tilde y}$ is sent to the channel decoder and the semantic decoder in turn for decoding, to generate the final image $\mathbf{\tilde x}$, which is modeled as

\begin{equation}
\label{7}
\begin{array}{l}
{\mathbf{x_{cd}}} = {f_{cd}}(\mathbf{\tilde y},\mathbf{{\theta _{cd}}})
\end{array}
\end{equation}
and
\begin{equation}
\label{8}
\begin{array}{l}
\mathbf{\tilde x} = {f_{sd}}(\mathbf{x_{cd}},\mathbf{\theta _{sd}}),
\end{array}
\end{equation}
where $\mathbf{x_{cd}}$ is the channel decode result. $\mathbf{\theta _{cd}}$ and $\mathbf{\theta _{sd}}$ are the model parameters of semantic encoder and channel encoder, respectively. 

In summary, the IRST framework emphasizes the coordination and feedback mechanism between modules. The semantic segmentation module uses the SME algorithm to enhance boundary recognition, providing more accurate semantic labels for subsequent semantic selection. The semantic selection module dynamically adjusts the importance distribution of image content based on task requirements and channel conditions. The data transmission module selects the optimal coding path based on real-time channel quality, achieving a balance between computational resources and transmission efficiency. These three components work together to build a complete communication link with semantic perception, task-driven and channel adaptive ability.

\section{Proposed Semantic Segmentation Module}
\label{section3}
In this section, we present the semantic segmentation module. Firstly, we introduce the architecture of the semantic segmentation model. Then, we propose an SME scheme. 

\subsection{Semantic Segmentation Module Architecture}

In satellite-ground communication, especially in remote sensing and satellite image analysis tasks, satellites face the challenge of data transmission capabilities. Due to the limited capacity of the satellite transmission link, they can not directly transmit all the collected data. Therefore, we adopt a selective transmission strategy to preferentially send the most valuable information.
In the designed image segmentation selection architecture, we first perform semantic segmentation on different types of target in the image. In this part, we use SegNet \cite{badrinarayanan2017segnet} as the main architecture. In this way, efficient feature encoding and decoding can be achieved, and spatial level information can be preserved while compressing features. Meanwhile, semantic segmentation is helpful to distinguish different types of task objectives, to evaluate the importance of each region and achieve priority transmission.
The semantic segmentation module consists of three main parts: encoder, decoder, and classification layer, with detailed descriptions as follows.

\subsubsection{Encoder}
The encoder uses the first 13 convolutional layers of Visual Geometry Group 16 (VGG16) for feature extraction. This step uses standard 2D convolution, which can be expressed as 

\begin{equation}
\label{9}
\begin{array}{l}
T(i,j) = \sum\limits_{m = 0}^{M - 1} {\sum\limits_{n = 0}^{N - 1} {S(i + m,j + n) \cdot W(m,n) + b} },
\end{array}
\end{equation}
where $S(i,j)$ is the input feature map, $W(m,n)$ is the convolution kernel, $b$ is the bias term, and $T(i,j)$ is the output feature map. Then, the resolution is reduced by a $2*2$ maximum pooling to reduce the amount of calculation. This step can be expressed as 

\begin{equation}
\label{10}
\begin{aligned}
T(i,j) = \max\big\{ 
& S(2i,2j),\,
S(2i + 1,2j),\,
S(2i,2j + 1),\\
& \hspace{40pt}
S(2i + 1,2j + 1)
\big\}
\end{aligned}
\end{equation}

At the same time, encoder stores pooled indexes for subsequent upsampling. 

\subsubsection{Decoder}
In the decoder stage, the model uses the pooling index to perform upsampling, which can recover the boundary information and improve the segmentation accuracy. This step can be expressed as

\begin{equation}
\label{11}
\begin{aligned}
S'(i,j) = \bigg\{ \begin{array}{*{20}{c}}
{S(i' ,j' ),}&{{\rm{ \hspace{5pt}pooled \hspace{5pt}index \hspace{5pt}position;}}}\\
{0,}&{{\rm{other \hspace{5pt}position.}}}
\end{array}
\end{aligned}
\end{equation}
Here, $S(i',j')$ represents the eigenvalue obtained from the encoder, and $S'(i,j)$ represents the upsampling result after reconstruction in the decoder. Compared with the traditional deconvolution or bilinear interpolation methods, this mechanism has stronger geometric structure retention ability and can complete the recovery of spatial resolution without introducing redundant information \cite{badrinarayanan2017segnet}.

\subsubsection{Classification Layer}
The classification layer uses Softmax to classify the pixels and output the category of each pixel, which is denoted as

\begin{equation}
\label{12}
\begin{array}{l}
P_c = \frac{e^{{\rm{z}}_c}}{\sum\nolimits_{k = 1}^K {e^{{z_k}}} },
\end{array}
\end{equation}
where $P_c$ is the probability that the pixel belongs to category $c$, $z_c$ is the score of category $c$, and $K$ is the total number of categories.

\subsection{Segmentation Enhancement Scheme}

As mentioned above, we use SegNet for semantic segmentation of input images. However, it may suffer from reduced spatial resolution during the encoding process, leading to imprecise object boundaries. Additionally, their pixel-level mapping strategy increases the likelihood of isolated noise points around edges, which undermines segmentation accuracy. To address this issue, we employ an SME scheme to replace individual pixels with dominant neighborhood color values, thereby improving regional consistency and enhancing classification precision. The detailed procedure of SME is described in \textbf{Algorithm \ref{alg1}}, with descriptions as follows.

First, a copy of the image $\mathbf{\tilde {z}}$ is created to ensure that the modification of the result does not affect the original image $\mathbf{z}$. Then, obtain the height $(h)$, width $(w)$, and number of channels (such as RGB channels, usually 3) of the image. Then, the three channels (R, G, B) of the image are packaged into an integer representation, which is convenient for subsequent operations, such as quickly comparing the colors of different pixels. After that, only the internal pixel area of the image (that is, from $(1, 1)$ to $(h-2, w-2)$) is processed to avoid cross-border errors. For each target pixel, take its neighborhood matrix $3*3$, then remove the center pixel value, and count the integer representation of each RGB color combination in the neighborhood to obtain its occurrence times. If a color value appears in the neighborhood more than or equal to $7$ times (i.e., dominant), the RGB value of the center pixel is replaced with the color value (the replacement process is completed by decoding the packaged integer color value back to the RGB channels). Finally, the modified image $\mathbf{\tilde {z}}$ is returned.

\begin{algorithm}[tbp]
%\small
\renewcommand{\algorithmicrequire}{\textbf{Input:}}
\renewcommand{\algorithmicensure}{\textbf{Output:}}
\caption{The Steps of Segmentation Model Enhancement}
\label{alg1}
\begin{algorithmic}[1]
\Require Input image $\mathbf{z}$, the dimension is $(h, w, c)$, where $h$ is the image height, $w$ is the width, and $c = 3$ represents the RGB channel.
\Ensure Modified image $\mathbf{\tilde {z}}$ , where the dimension is the same as $\mathbf{z}$.
\State Create a copy of the result $\mathbf{\tilde {z}} = copy (\mathbf{z})$;
\State Get height and width $h, w$;
\State Shift the red $(R)$, green $(G)$ and blue $(B)$ channels and added respectively;
\For {$i$ in range $(1, h-1)$}
   \For {$j$ in range $(1, w-1)$}
        \State Get the $3*3$ neighborhood of the target pixel and \\ \hspace{25pt} remove the central pixel;
        \State Count the number of occurrences of each color in \\ \hspace{25pt} the neighborhood;
        \If {The number of occurrences of a color value \\ \hspace{25pt} $max_{count}  >= 7$}
             \State Replace the center pixel value;
        \EndIf
   \EndFor     
\EndFor
\State Return the modified image $\mathbf{\tilde {z}}$.
\end{algorithmic}
\end{algorithm}

Then, let us explain the SME algorithm from a theoretical point of view. Let $z(i,j)$ denote the pixel at position $(i,j)$, and $\mathcal{N}(i,j)$ denote its $3 \times 3$ neighborhood excluding the center. Define the majority function as 

\begin{equation}
\label{13}
\begin{array}{l}
M(i,j) = \arg\max_{c \in \mathcal{C}} \; \text{count}_{\mathcal{N}(i,j)}(c),
\end{array}
\end{equation}
where $\mathcal{C}$ is the set of color values. The SME update rule is modeled as

\begin{equation}
\label{14}
\begin{array}{l}
z'(i,j) =
\begin{cases}
M(i,j), & \text{if } \max_c \text{count}_{\mathcal{N}(i,j)}(c) \geq \tau, \\
z(i,j), & \text{otherwise},
\end{cases}
\end{array}
\end{equation}
with threshold \(\tau = 7\). 

This mechanism can be interpreted as a local denoising operator that minimizes the probability of edge misclassification. Specifically, if the probability of noise corruption per pixel is $p$, the probability that a majority of neighbors are corrupted is bounded by 

\begin{equation}
\label{15}
\begin{array}{l}
{P_{error}} \le \sum\limits_{k = \tau }^8 {\left( \begin{array}{c}
8\\
k
\end{array} \right)} {p^k}{(1 - p)^{8 - k}}.
\end{array}
\end{equation}
with threshold $\tau = 7$. 

Thus, SME significantly reduces edge error when $p < 0.3$, ensuring the accuracy in semantic segmentation.

\section{Proposed Task-oriented Semantic Selection Module}
\label{section4}
In this section, we present the semantic selection module based on the task requirement. Firstly, we introduce the transmission effect evaluation scheme. Then, we give the specific operation process for semantic selection.

\subsection{Transmission Effect Evaluation Scheme}

In the process of satellite image transmission, the communication bandwidth between the satellite and the ground is often limited, which cannot support the complete transmission of all image data \cite{araniti2015effective}. The amount of high-definition satellite image data is huge, and the transmission link may not be able to carry all the information. Therefore, it is necessary to compress, filter, or extract key areas of the image to reduce the amount of data while ensuring effective transmission of the core information \cite{schwartz2023satellite}. Meanwhile, the satellite operates in the space environment, which is affected by electromagnetic interference, atmospheric influence, equipment noise, and other factors, resulting in the transmission image may contain a lot of noise \cite{kaur2011electromagnetic}. If the unprocessed image is transmitted directly, it can affect the accuracy of ground data analysis \cite{araniti2015effective}. Therefore, by selecting and extracting the transmission content, the satellite system can improve the transmission efficiency of key data under limited bandwidth conditions and reduce the impact of noise on image quality, making ground analysis more reliable.

To achieve this goal, we construct a transmission effect evaluation function. The proposed function offers a systematic and objective methodology for evaluating the transmission quality of satellite imagery under bandwidth constraints and noise interference, thereby avoiding reliance on subjective visual assessments. Furthermore, it facilitates the identification and prioritization of essential image content, enabling the satellite system to transmit high-value data while maximizing the information yield within limited communication bandwidth.
The proposed transmission effect evaluation function is expressed as

\begin{equation}
\label{16}
\begin{array}{l}
{\hat y} = f(X_s,X_c),
\end{array}
\end{equation}
where $\hat y$ is the result of the transmission effect evaluation, $X_{s}$ represents the image or the type of task, and $X_c$ represents the external conditions (such as bandwidth, SNR, etc.). The goal of the model is to find the optimal $f( \cdot )$ such that the estimated value $\hat y$ is close to the true value $y$.

In order to better capture the nonlinear relationship between task complexity, bandwidth, SNR and other factors, and improve the prediction accuracy, we use linear regression to model. Firstly, we establish a linear regression model as

\begin{equation}
\label{17}
\begin{array}{l}
\hat{y} = w_0 + w_1 X_s + w_2 X_c + \epsilon,
\end{array}
\end{equation}
where $N$ is the number of samples, ($w_1$, $w_2$) is the parameter to optimize the model and $\epsilon$ is the noise error. Then, we use the mean square error (MSE) as the loss function, which is expressed as

\begin{equation}
\label{18}
\begin{array}{l}
L(w) = \frac{1}{N}\sum\limits_{i = 1}^N {{{({y_i} - {{\hat y}_i})}^2}}.
\end{array}
\end{equation}
For the parameters $w$, the loss function gradient is 

\begin{equation}
\label{19}
\begin{array}{l}
\frac{{\partial L}}{{\partial w}} =  - \frac{2}{N}\sum\limits_{i = 1}^N {{x_i}({y_i} - {{\hat y}_i})},
\end{array}
\end{equation}
where $x_i$ is the input feature of the corresponding sample. Finally, we use the gradient descent algorithm to update the weights as
\begin{equation}
\label{20}
\begin{array}{l}
w = w - \alpha \frac{{\partial L}}{{\partial w}},
\end{array}
\end{equation}
where $\alpha$ is the learning rate.

Through the evaluation scheme, when the new image data are input, the image features of the image are first extracted, and then the current channel conditions are input. Based on these factors, the scheme will evaluate the image recovery result $\hat y$ at the end of the output. If $\hat y$ is too low, the image resolution is reduced and the amount of data is reduced. At the same time, priority is given to the transmission of key areas to improve the effective information ratio.

\subsection{Theoretical Modeling of Semantic Selection}

To enhance the theoretical rigor of the semantic selection module, we formalize the selection process as a task-driven optimization problem under channel constraints. Let the original image be denoted by $\mathbf{s}\in \mathbb{R}^{H \times W \times C}$, and the semantic segmentation output by $\mathbf{s_{ss}} \in \mathbb{R}^{H \times W \times K}$, where $K$ is the number of semantic categories. The task requirement is represented by a weight vector $q \in \mathbb{R}^K$, and the current channel condition (e.g. SNR) is denoted by $\gamma \in \mathbb{R}$.

We define the semantic relevance score $\mathcal{I}(i,j)$ for each pixel location $(i,j)$ as: 
\begin{equation}
\label{21}
\mathcal{I}(i,j) = \sum_{k=1}^{K} \mathbf{s_{ss}}(i,j,k) \cdot q_k,
\end{equation}
where $\mathbf{s_{ss}}(i,j,k)$ is the probability that pixel $(i,j)$ belongs to category $k$, and $q_k$ reflects the task-specific importance of category $k$.

Given the limited transmission budget $B(\gamma)$, which is a function of the channel condition $\gamma$, the selection of image regions $x \subset s$ must satisfy:
\begin{equation}
\label{22}
\sum_{(i,j) \in x} \mathcal{C}(i,j) \leq B(\gamma),
\end{equation}
where $\mathcal{C}(i,j)$ denotes the transmission cost of pixel $(i,j)$, approximated by its encoding complexity or bit-length.

The semantic selection task is thus cast as the following constrained optimization problem:
\begin{equation}
\label{23}
\max_{x \subset s} \sum_{(i,j) \in x} \mathcal{I}(i,j) \quad \text{s.t.} \quad \sum_{(i,j) \in x} \mathcal{C}(i,j) \leq B(\gamma).
\end{equation}

To reduce computational overhead and enable real-time deployment on resource-constrained satellite platforms, we propose a lightweight scoring function $\mathcal{S}(i,j)$ that incorporates channel awareness:
\begin{equation}
\label{24}
\mathcal{S}(i,j) = \mathcal{I}(i,j) \cdot \phi(\gamma),
\end{equation}
where $\phi(\gamma)$ is a channel-adaptive modulation function. For instance, a threshold-based formulation may be adopted:
\begin{equation}
\label{25}
\phi(\gamma) = 
\begin{cases}
1, & \gamma \geq \gamma_{th}, \\
\alpha, & \gamma < \gamma_{th},
\end{cases}
\end{equation}
{with $\alpha < 1$ serving to suppress low-importance regions under poor channel conditions.

Based on $\mathcal{S}(i,j)$, the image is selectively compressed via a spatial filtering function $\mathcal{B}(s; \mathcal{S}, \tau)$, where $\tau$ is a tunable threshold. Pixels with $\mathcal{S}(i,j) < \tau$ are blurred or downsampled, while high-score regions are preserved in full fidelity.

This theoretical framework enables the semantic selection module to dynamically balance task relevance and transmission efficiency, while maintaining computational tractability. It also provides a principled basis for future extensions such as multi-task prioritization and reinforcement-based selection strategies.

\subsection{Semantic Selection Process}

As discussed earlier, the satellite-ground communication system is inherently constrained by limited bandwidth, due to the size and computational capacity of individual satellites \cite{araniti2015effective}. Furthermore, the space environment introduces significant noise into the transmission process (including electromagnetic interference, atmospheric disturbances, and equipment-related factors), which adversely affects the quality of image reconstruction. To address these challenges associated with satellite data transmission under constrained bandwidth and unstable channel conditions, we propose a semantic selection strategy guided by a transmission effect evaluation scheme. The strategy selects the appropriate content for transmission by analyzing the transmission task and the current channel state, and maximizes the use of bandwidth resources while ensuring the completion of the task. The detailed procedural flow of this method is presented in \textbf{Algorithm}~\ref{alg2}.

\begin{algorithm}[tbp]
%\small
\renewcommand{\algorithmicrequire}{\textbf{Input:}}
\renewcommand{\algorithmicensure}{\textbf{Output:}}
\caption{The Steps of Task-oriented Semantic Selection}
\label{alg2}
\begin{algorithmic}[1]
\Require Target image $\mathbf{s}$, pre-trained segmentation model $f_s$, pre-trained evaluation model $f_e$, color mapping table $C$, SNR $\gamma$.
\Ensure Optimized image $\mathbf{x}$ after processing, where the dimension is the same as $\mathbf{s}$.
\State Obtain the input value $\mathbf{\hat{s}}$ of the image $\mathbf{s}$ after loading and preprocessing;
\State Obtain the classification result matrix $\mathbf{O}$ of $\mathbf{\hat{s}}$ by using the segmentation model $f_s$; 
\State Calculate the category index, and use the color mapping table $C$ to color, generate color segmentation map $\mathbf{O_{c}}$;
\State Generate a background image $\mathbf{O_{b}}$ according to the target category;
\State Get the transmission effect evaluation result $\hat y$ in current $\gamma$ by using evaluation model $f_e$;
\State Perform the mean fuzzy filtering of $\mathbf{O_{b}}$ according to the transmission effect evaluation result $\hat y$.
\State Fuse the fuzzy filtered image and the original mask to generate the final result $\mathbf{x}$.
\State Return the modified image $\mathbf{x}$.
\end{algorithmic}
\end{algorithm}

Firstly, the semantic segmentation model is used to process the input image, identify different categories, and apply color mapping. Then, according to the different tasks, the target area of the task is screened by mask operation to generate an adjusted binary image. On this basis, the transmission effect evaluation model is used to evaluate the recovery of the image after channel transmission in the current noise environment. In this scheme, we use the pilot-based minimum mean square error (MMSE) channel estimation method to recover the channel response, calculate the instantaneous SNR at the receiver, and return the value to the satellite.

According to the evaluation results, different mean fuzzy filtering schemes are used to adaptively adjust the non-main part of the image to enhance the visual effect and improve the accuracy of the analysis. Finally, in combination with mask correction technology, an optimized image is generated to prepare for subsequent data transmission. Figure \ref{fig3} shows the effect. In this way, according to the different transmission tasks, the proposed solution can reduce 10\%-80\% of the transmission content and greatly save bandwidth.

\begin{figure*}[!t]
\centerline{\includegraphics[width=0.8\textwidth]{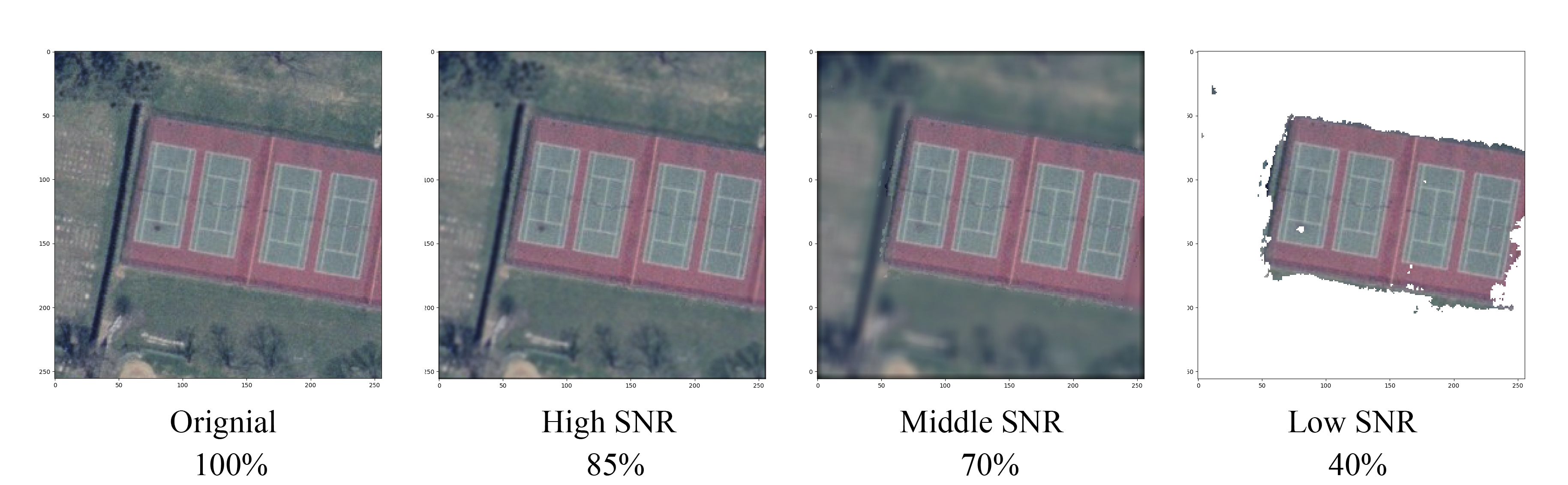}}
\caption{The fuzzy filtering effect under different SNR conditions, the lower the SNR, the stronger the fuzzy filtering. When the SNR is at the minimum, only the part of the task requirement is transmitted.}
\label{fig3}
\end{figure*}

\section{Proposed Dynamic SNR-based Transmission Module}
\label{section5}
In this section, we introduce the data transmission module of the model under dynamic SNR. Firstly, we introduce the semantic codec module. Then, we propose a codec module based on the stacked channel. Finally, we give the module training scheme.

\subsection{Semantic Codec Architecture}
In our proposed image transmission framework, the semantic codec module adopts the Swin Transformer model. Swin Transformer is a visual Transformer model designed for computer vision tasks, which is proposed by the Microsoft research team \cite{liu2021swin}. It uses a layered architecture and a shifted windows mechanism to achieve a balance between computational complexity and performance, making it suitable for tasks such as image classification, target detection, and semantic segmentation. This module is composed of an encoder and a decoder. 

In the encoder part, the role of the encoder is to extract multi-scale features, which is mainly composed of Swin Transformer blocks and patch merging. The Swin Transformer blocks part adopts the Window-based Multi-head Self-Attention (W-MSA) mechanism and the Shifted Windows Multi-head Self-Attention (SW-MSA) mechanism \cite{liu2021swin}. W-MSA is used to calculate self-attention in a local window to improve computational efficiency. SW-MSA realizes the interaction of information across windows through the window shift mechanism. The patch merging part adopts the gradual merging of adjacent patches to improve the feature expression ability by reducing the spatial dimension. At the same time, this part adopts a hierarchical structure, so that the model can learn features of different scales.

For the decoder, the role of this part is to restore the spatial resolution of the image and generate the final output. The decoder mainly includes three parts: reverse patch merging, Swin Transformer blocks and final output layer. Reverse patch merging is used to gradually restore the spatial structure of the image. Swin Transformer blocks continues to use the self-attention mechanism for feature fusion and refinement. The final output layer is used to generate the final results and restore the original image.

By establishing the semantic codec framework, not only can the bit information be accurately transmitted, but also the core meaning of the information can be extracted to ensure that the receiver can correctly understand the meaning of the information, thereby optimizing communication efficiency and reducing redundant data transmission.

\subsection{Dynamic SNR Channel Codec Architecture}

Due to atmospheric disturbances (such as precipitation, clouds, and water vapor), terrain occlusion, and satellite orbital variations, the SNR of satellite-ground transmission fluctuates significantly over time \cite{toka2024ris}. These fluctuations lead to frequent data loss and elevated bit error rates, particularly under low-SNR conditions. Given the limitations in onboard capacity and power consumption, satellites are unable to adapt to each channel state individually. Therefore, we propose a dynamic SNR-adaptive codec architecture based on the concept of stacking \cite{kim2022stacked}. This three-tier design enables flexible channel coding and decoding across varying SNR levels: it activates only the first layer under high-SNR conditions, while utilizing all three layers when the SNR deteriorates. Figure \ref{fig4} illustrates the architecture in detail.

\begin{figure}[!t]
\centerline{\includegraphics[width=0.48\textwidth]{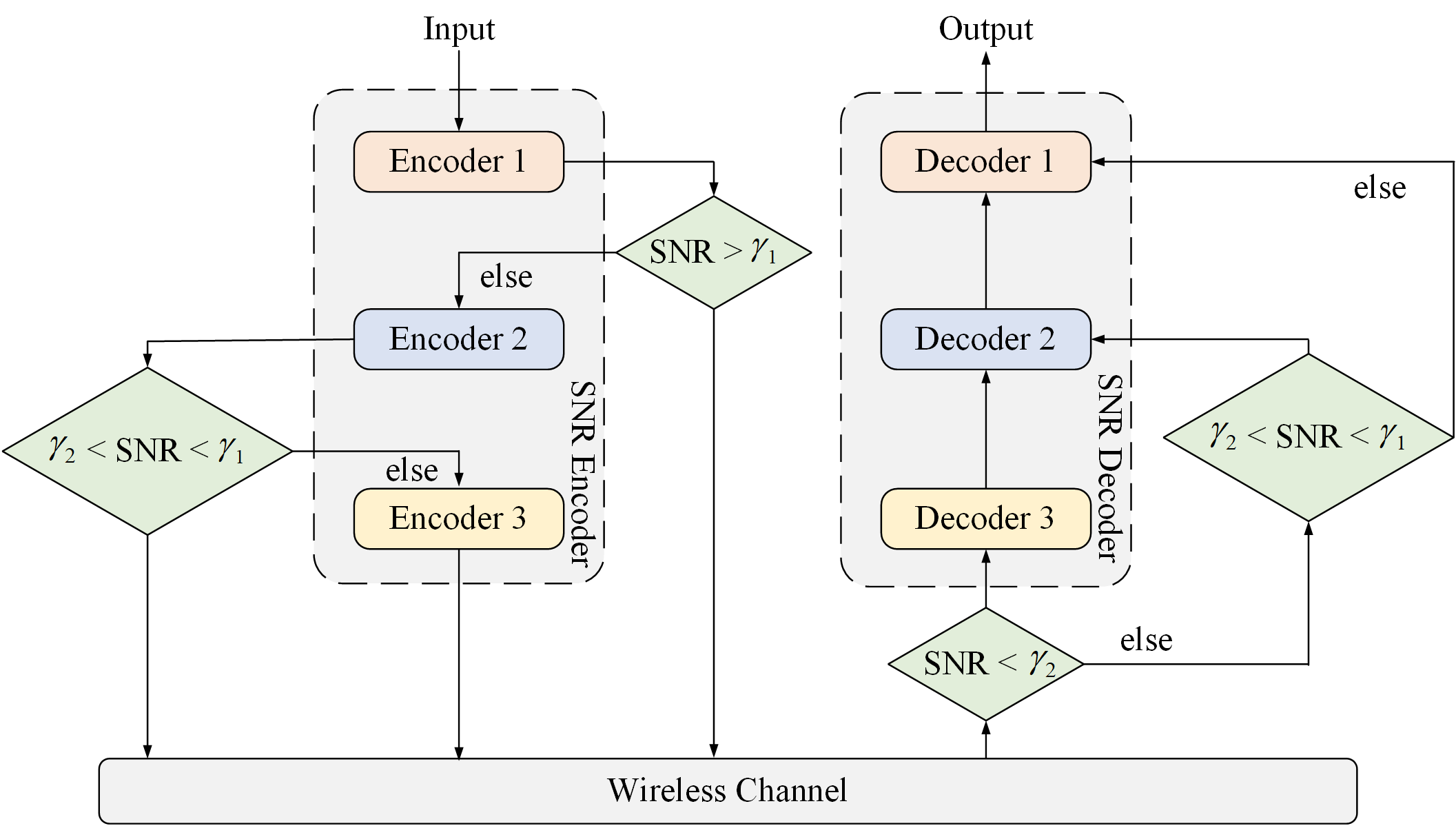}}
\caption{The hierarchical architecture of SNR codec, which consists of three layers of codec. Different layers of codec are chosen according to the level of SNR. All codecs are used when SNR is the lowest.}
\label{fig4}
\end{figure}

The model consists of two core modules: SNR selective encoder and SNR selective decoder. The encoder uses a set of $3*3$ convolution layers to process the input $\mathbf{u}$, and selectively applies feature extraction at different depths according to SNR. The decoder uses reverse convolution for signal recovery and also dynamically adjusts the network structure according to the SNR.

To formalize the adaptive transmission process under dynamic SNR conditions, we define the end-to-end transmission pipeline as a composition of semantic encoding, channel encoding, channel fading, and decoding operations.

The encoder is segmented into $L$ layers, each with activation condition $\gamma_l$ based on the current SNR $\gamma$. The activation function is defined as:
\begin{equation}
\label{26}
\delta_l(\gamma) = 
\begin{cases}
1, & \gamma < \gamma_l, \\
0, & \gamma \geq \gamma_l,
\end{cases}
\end{equation}
where $\delta_l(\gamma)$ indicates whether layer $l$ is activated under SNR $\gamma$.

The overall quality of transmission is evaluated using a semantic fidelity metric $\mathcal{F}(\mathbf{\tilde x}, \mathbf{x})$, which may include perceptual similarity and task-specific precision. The transmission optimization objective is as follows:
\begin{equation}
\label{27}
\max_{\theta_{se}, \theta_{ce}} \mathbb{E}_{\gamma, h, n} \left[ \mathcal{F}(\mathbf{\tilde x}, \mathbf{x}) \right] \quad \text{s.t.} \quad \mathcal{C}(y) \leq B(\gamma),
\end{equation}
where $\mathcal{C}(y)$ denotes the transmission cost and $B(\gamma)$ is the bandwidth budget under SNR $\gamma$.

Through this architecture, it can simplify the calculation and improve efficiency at high SNR, enhance feature extraction at low SNR, improve robustness, and finally optimize signal transmission and recovery quality. This method combines SemCom and adaptive signal processing and has excellent performance in communication environments with limited bandwidth or severe noise.

\subsection{The Training Scheme of SNR Channel Codec}

To address the challenges posed by varying SNR conditions, we propose a dynamic SNR channel codec employing a three-tier architecture in the previous subsection. However, independently training the model across three separate SNR intervals may result in parameter mismatches between adjacent layers due to inconsistencies in training environments. Since the output of one layer directly serves as the input for the next, these mismatches can adversely affect reconstruction quality. To mitigate this issue, we introduce an SNR channel parameter training scheme. This scheme utilizes a progressive, layer-wise training approach and incorporates distinct SNR regions to ensure effective adaptation to dynamic channel conditions. As a result, the model achieves tailored encoding strategies for different SNR environments while maintaining a unified parameter set, thereby significantly reducing memory usage without compromising coding performance. The detailed procedure is presented in \textbf{Algorithm}~\ref{alg3}.

\begin{algorithm}[tbp]
%\small
\renewcommand{\algorithmicrequire}{\textbf{Input:}}
\renewcommand{\algorithmicensure}{\textbf{Output:}}
\caption{The Training Steps of the SNR Channel Codec}
\label{alg3}
\begin{algorithmic}[1]
\Require Encoder architecture $e_1$, $e_2$, $e_3$; decoder architecture $d_1$, $d_2$, $d_3$; dynamic SNR range $\gamma_{min}$, $\gamma_{max}$; SNR switching threshold $\gamma_1$, $\gamma_2$; train epoch $E_p$; input data $\mathbf{x_{in}}$.
\Ensure Encoder parameters $\mathbf{\theta_{e1}}$, $\mathbf{\theta_{e2}}$, $\mathbf{\theta_{e3}}$; decoder parameters $\mathbf{\theta_{d1}}$, $\mathbf{\theta_{d2}}$, $\mathbf{\theta_{d3}}$.
\State Add encoder1 architecture $e_1$ and decoder1 architecture $d_1$;
\For {epoch = 1 : $E_p$}
     \State Randomly generate $\gamma_i$ satisfying $\gamma_i \in (\gamma_{min},\gamma_{max})$;
     \State Calculate the output $\mathbf{x_{out1}}$ of $\mathbf{x_{in}}$ through $e_1$,  \\ \hspace{15pt}wireless channel and $d_1$;
     \State Update $\mathbf{\theta_{e1}}$ and $\mathbf{\theta_{d1}}$ through calculating the loss \\ \hspace{15pt}function of $\mathbf{x_{out1}}$ and $\mathbf{x_{in}}$;
\EndFor
\State Add encoder2 architecture $e_2$ and decoder2 architecture $d_2$;
\State Freeze  $\mathbf{\theta_{e1}}$ and $\mathbf{\theta_{d1}}$;
\For {epoch = 1 : $E_p$}
     \State Randomly generate $\gamma_i$ satisfying $\gamma_i \in (\gamma_{min},\gamma_1)$;
     \State Calculate the output $\mathbf{x_{out2}}$ of $\mathbf{x}$ through $e_1$, $e_2$, \\ \hspace{15pt}wireless channel, $d_2$ and $d_1$;
     \State Update $\mathbf{\theta_{e2}}$ and $\mathbf{\theta_{d2}}$ through calculating the loss \\ \hspace{15pt}function of $\mathbf{x_{out2}}$ and $\mathbf{x_{in}}$.
\EndFor
\State Add encoder3 architecture $e_3$ and decoder3 architecture $d_3$;
\State Freeze $\mathbf{\theta_{e2}}$ and $\mathbf{\theta_{d2}}$;
\For {epoch = 1 : $E_p$}
     \State Randomly generate $\gamma_i$ satisfying $\gamma_i \in (\gamma_{min},\gamma_2)$;
     \State Calculate the output $\mathbf{x_{out3}}$ of $\mathbf{x_{in}}$ through $e_1$, $e_2$, \\ \hspace{15pt}$e_3$, wireless channel, $d_3$, $d_2$ and $d_1$;
     \State Update $\mathbf{\theta_{e3}}$ and $\mathbf{\theta_{d3}}$ through calculating the loss \\ \hspace{15pt}function of $\mathbf{x_{out3}}$ and $\mathbf{x_{in}}$.
\EndFor
\end{algorithmic}
\end{algorithm}

Firstly, the input data $x$ is trained using the initial encoder $e_1$ and decoder $d_1$, and the parameters $\theta_1$ and $\theta'_1$ are optimized based on the loss function. Subsequently, additional encoder $e_2$ and decoder $d_2$ are added and further optimized for lower SNR on the basis of freezing previous training parameters. The process is progressively extended to the third set of encoder $e_3$ and decoder $d_3$, enabling the model to handle a wider range of channel conditions. Finally, the scheme outputs a series of optimized encoder parameters $\theta_1$, $\theta_2$, $\theta_3$ and decoder parameters $\theta'_1$, $\theta'_2$, $\theta'_3$ to improve the overall robustness and adaptability of the communication system.

\section{Simulation Results and Analysis}
\label{section6}
\subsection{Theoretical Complexity Analysis}
To rigorously assess the computational efficiency of the proposed IRST framework, we present a theoretical analysis of the time complexity associated with its three principal modules: the semantic segmentation module, the semantic selection module, and the data transmission module. This analysis elucidates the computational demands of each component under varying image resolutions and channel conditions, thereby establishing a theoretical foundation for subsequent runtime benchmarking.

\subsubsection{Semantic Segmentation Module}
The semantic segmentation module adopts the SegNet architecture, and its encoder part is based on the first 13 layers of VGG16 convolutional network. The time complexity of each layer convolution operation is $O(H \cdot W \cdot {K^2} \cdot {C_{in}} \cdot {C_{out}})$. Among them, $H$,$W$ are the size of the feature map, $K$ is the size of the convolution kernel, and $C_{in}$, $C_{out}$ are the size of the convolution kernel. The decoder uses upsampling based on pooling index, and the complexity is $O(H \cdot W)$. The classification layer uses Softmax to classify each pixel with a complexity of $O(H \cdot W \cdot C)$, where $C$ is the number of categories. Therefore, the overall complexity of the semantic segmentation module is approximately $O(H \cdot W \cdot C^2)$.

\subsubsection{Semantic Selection Module}
The module evaluates and blurs the image region according to the task requirements and the current SNR. Due to the use of a lightweight pixel neighborhood statistical method, the complexity mainly comes from the local window operation for each pixel (such as 3 × 3 neighborhood), so the overall complexity is $O(H \cdot W )$.

\subsubsection{Data Transmission Module}
The module includes Swin Transformer semantic encoder and SNR segmented channel encoder. The window attention mechanism of Swin Transformer reduces the complexity from $O(N^2 \cdot d)$ of traditional Transformer to $O(N \cdot d \cdot w^2)$. Among them, $N = H \cdot W$, $d$ is the feature dimension, and $w$ is the window size. The channel encoder dynamically activates sub-networks of different depths according to SNR conditions. In the worst case, the complexity is $O(H \cdot W \cdot D)$, where $D$ is the maximum coding depth. Under high SNR conditions, some paths can be skipped to further reduce the computational load.

In summary, the IRST framework demonstrates effective control over computational complexity through modular architecture and conditional activation strategies. This design ensures robust semantic representation while offering adaptability to deployment constraints and runtime efficiency.
\subsection{Dataset and Simulation Parameters}
We use the Dense Labeling Remote Sensing Dataset (DLRSD) \cite{shao2018performance} for image visualization training and verification. DLRSD is a high-density labeled remote sensing image dataset, which is mainly used for multi-label tasks, such as remote sensing image retrieval, classification, and semantic segmentation. It contains 2100 remote sensing images, each with a size of 256 × 256 pixels and a resolution of 2 m. It uses GaoFen-1 and ZiYuan-3 satellite data, covering 17 main land cover categories, including aircraft, bare soil, buildings, cars, grasslands, oceans, ships, etc.
To further validate the generalizability of the proposed model, we incorporate the NWPU VHR-10 dataset \cite{cheng2014multi} as an auxiliary benchmark. It is a widely recognized and challenging dataset in the domain of remote sensing image target detection, comprising 800 high-resolution optical remote sensing images. It is extensively utilized in tasks such as object detection and geospatial object recognition, making it a robust testbed for evaluating model performance in real-world scenarios.

We utilize the light SR channel in our implementation. For the IRST model, the simulation channel uses the threshold of the SNR codec we set as $\gamma_1=3$ dB, $\gamma_2=-3$ dB. The main parameters are shown in Table \ref{Table.2} and Table \ref{Table.3}. All configurations are implemented using the PyTorch framework and are executed on a NVIDIA A10 GPU.

\begin{table}[htbp]
\centering
\caption{Simulation Parameters}
\begin{tabular}{|c|c|}
 \hline
Parameter Name & Value  \\ \hline
Average Power of  Sattered Component $b_0$ & 0.158   \\ \hline
Nakagami-$m$  & 19.4   \\ \hline
 LOS Component $\Omega$ & 1.29  \\ \hline
Batch Size      & 8         \\ \hline
Optimizer      & Adam          \\ \hline
Loss Function  & MSE          \\ \hline
Learning Rate  & 0.0001        \\ \hline
Train Epoch    & 100          \\ \hline
SNR Range      & -10dB - 10dB  \\ \hline
Activation Function      & ReLU   \\ \hline
Weight Initialization      & Xavier  \\ \hline
Regularization Coefficient      & 0.01  \\ \hline
\end{tabular}
\label{Table.2}
\end{table}

\begin{table}[htbp]
\caption{Main Structure of the Model}
\begin{tabular}{ccc}
\hline
                      & \textbf{Layers}                    & \textbf{Output Size}   \\
                      \hline
 & Conv2d+ReLU  x  3         & (1,256,16,64) \\
 \textbf{Semantic Segmentation}                     & ConvTranspose2d+ReLU  x 2 & (1,64,64,64)  \\
                      & Conv2d                    & (1,3,64,64)   \\\hline
      & PatchEmbed                & 256           \\
                      & SwinTransformerBlock x 2  & 128           \\
 \textbf{Semantic Coding}                      & SwinTransformerBlock x 4  & 256           \\
                      & LayerNorm                 & 256           \\
                      & Linear                    & 96            \\\hline
       & Encoder1                  & (1,64,64,64)  \\
 \textbf{Channel Coding}                      & Encoder2                  & (1,128,64,64) \\
                      & Encoder3                  & (1,256,64,64) \\\hline
      & Decoder3                  & (1,128,64,64) \\
    \textbf{Channel Decoding}                  & Decoder2                  & (1,64,64,64)  \\
                      & Decoder1                  & (1,1,64,64)   \\\hline
   & SwinTransformerBlock x 4  & 256           \\
    \textbf{Semantic Decoding}                    & SwinTransformerBlock x 2  & 128           \\
                      & PatchReverseMerging       & 128           \\
                      & Linear                    & 256           \\\hline          
\end{tabular}
\label{Table.3}
\end{table}

\subsection{Comparison Schemes}
\subsubsection{Deep JSCC}
Deep JSCC \cite{bourtsoulatze2019deep} adopts an end-to-end deep neural network framework that directly maps image pixels to channel input symbols, effectively bypassing the conventional processes of image compression and channel coding. Typically structured as an autoencoder, it comprises a Convolutional Neural Network (CNN)-based encoder and decoder, with a non-trainable channel simulation module—commonly modeled as an Additive White Gaussian Noise (AWGN) or Rayleigh fading channel—embedded between them.

\subsubsection{WITT}
WITT \cite{yang2024swinjscc} adopts Swin Transformer to extract the hierarchical semantic features of the image, and performs feature compression and mapping through the attention mechanism, thereby enhancing the modeling ability of the image structure. The encoder converts the image into an embedded vector, and then performs channel mapping through a multi-head attention module; the decoder recovers the image from the received signal and maintains high fidelity.

\subsubsection{SS-DDPM}
SS-DDPM \cite{okhotin2023star} introduces a non-Gaussian diffusion framework that replaces the conventional Markov chain structure with a star-shaped diffusion process. It directly samples multiple noise states from the original data distribution, bypassing the need to define stepwise transition probabilities. 

\subsection{Performance Metrics}
\subsubsection{PSNR}
Peak Signal-to-Noise Ratio (PSNR) is an indicator used to measure the quality of an image or signal. By comparing the pixel difference between the original image and the transmitted image, the distortion degree of the image is quantified. PSNR is mainly calculated based on the mean square error at the pixel level. Because of its simplicity and objectivity, it is widely used in image compression, denoising, and quality monitoring during transmission \cite{yuan2024deep,bourtsoulatze2019deep,yang2024swinjscc}.
The higher the PSNR value, the higher the similarity between the two images, the better the quality. It is calculated as
\begin{equation}
\label{28}
PSNR = 10 \cdot lo{g_{10}}\left(\frac{{ma{x^2}}}{{MSE}}\right),
\end{equation}
where $max$ represents the maximum possible value of the pixel value, usually 255 in image processing, and $MSE$ represents the mean square error.

\subsubsection{SSIM}
Structural Similarity Index Measure (SSIM) is a perceptual metric developed to approximate the characteristics of human visual perception when assessing the quality of digital images. Unlike traditional pixel-based methods, SSIM evaluates image similarity from three aspects : brightness, contrast and structure, which is closer to the human visual system. In the process of image transmission, SSIM can more accurately reflect whether the image looks the same, even if the pixels are slightly different. This comprehensive approach enables SSIM to more accurately reflect perceived visual quality, particularly in scenarios where structural fidelity is crucial, such as image enhancement, super-resolution reconstruction, and facial image processing \cite{yang2024swinjscc,sagduyu2024joint,wu2024ccdm}. The SSIM index ranges from 0 to 1, with values closer to 1 indicating a higher degree of similarity between the compared images. It is calculated as

\begin{equation}
\label{29}
SSIM(x, y) = \frac{(2\mu_x\mu_y + c_1)(2\sigma_{xy} + c_2)}{(\mu_x^2 + \mu_y^2 + c_1)(\sigma_x^2 + \sigma_y^2 + c_2)},
\end{equation}

\noindent where $\mu_x$ and $\mu_y$ are the average values of $x$ and $y$, $\sigma_x^2$ and $\sigma_y^2$ are the variances of $x$ and $y$, and $\sigma_{xy}$ is the covariance of $x$ and $y$. $c_1$ and $c_2$ are constants used to maintain stability, usually set to $c_1 = (k_1L)^2$ and $c_2 = (k_2L)^2$, where $L$ is the dynamic range of pixel values, $k_1$ and $k_2$ are default values, generally take $k_1$ = 0.01 and $k_2$ = 0.03.

\subsubsection{LPIPS}
Learned Perceptual Image Patch Similarity (LPIPS) is a perceptual metric designed to evaluate the visual similarity between two images from the perspective of human perception. Unlike traditional pixel-wise metrics such as PSNR or SSIM, LPIPS compares deep feature representations extracted from pretrained neural networks, thereby capturing high-level semantic differences that are more aligned with human judgment. Due to its ability to reflect perceptual quality rather than mere pixel fidelity, LPIPS has been widely adopted in tasks such as image generation, super-resolution, style transfer, and adversarial robustness evaluation \cite{fu2024multimodal, guo2024a}. A lower LPIPS score indicates higher perceptual similarity between the reference and the distorted image. The LPIPS metric is formally calculated as

\begin{equation} 
\label{30} 
LPIPS(x, x_0) = \sum_l w_l \cdot \| f_l(x) - f_l(x_0) \|_2^2, \end{equation} 
where $f_l( \cdot )]$denotes the feature map extracted from layer $l$ of a pre-trained network, and $w_l$ represents the learned weight for each layer. $x$ and $x_0$ are the input and reference images, respectively.

\subsection{Results Analysis}
\subsubsection{Module Performance Verification}
Ablation experiments are employed to verify the performance improvement of the designed module.

Table \ref{Table.4} presents the comparative analysis of whether to use SME module, including accuracy, PSNR, SSIM and boundary Intersection over Union (IoU). The results indicate that although SME does not significantly improve overall accuracy, it significantly improves the quality of boundary segmentation. This is because the encoder part of SegNet uses a multi-layer maximum pooling operation. Although this enhances the translation invariance and robustness of the model, it also significantly reduces the spatial resolution of the feature map, making the boundary details blurred or lost during the downsampling process, resulting in a decrease in recognition rate. The SME improves the accuracy of location recognition by supplementing the points of the surrounding environment.

\begin{table}[]
\centering
\caption{SME Effect Comparison}
\begin{tabular}{|c|c|c|}
\hline
           & With SME     & Without SME  \\ \hline
Accuracy   & 0.728        & 0.714        \\ \hline
{PSNR} & {33.954} & {33.257} \\ \hline
{SSIM} & {0.558}  & {0.476}  \\ \hline
{Boundary IoU} & {0.718}  & {0.638}  \\ \hline
\end{tabular}
\label{Table.4}
\end{table}

Figure \ref{fig5} presents a performance comparison of remote sensing image transmission under conditions with and without the application of the proposed fuzzy filtering scheme. This scheme is designed to optimize bandwidth usage while preserving the visual integrity of task-critical regions. In Figure \ref{fig5a}, the transmitted image size is 160 KB, yielding an overall PSNR of 38.648 dB and a task-region PSNR of 39.077 dB. Although the image quality remains uniformly high, the substantial data size results in elevated transmission costs. In contrast, Figure \ref{fig5b} demonstrates the effectiveness of fuzzy filtering, reducing the image size to 117 KB—a compression rate of approximately 26.9 \%. While the overall SNR decreases to 32.747 dB and background details exhibit visible degradation, the task-region PSNR remains stable at 39.067 dB, closely matching the unfiltered result. This outcome suggests that the scheme maintains the perceptual quality of the task-critical area while significantly improving transmission efficiency. The design encapsulates the core principle of task-oriented SemCom: selectively suppressing redundant visual information to ensure semantic goal completion under constrained bandwidth conditions.

\begin{figure}[tbp]
  \centering
  \subfigure[Fuzzy filtering is not employed.]{
  \label{fig5a}
  \includegraphics[width=0.5\textwidth]{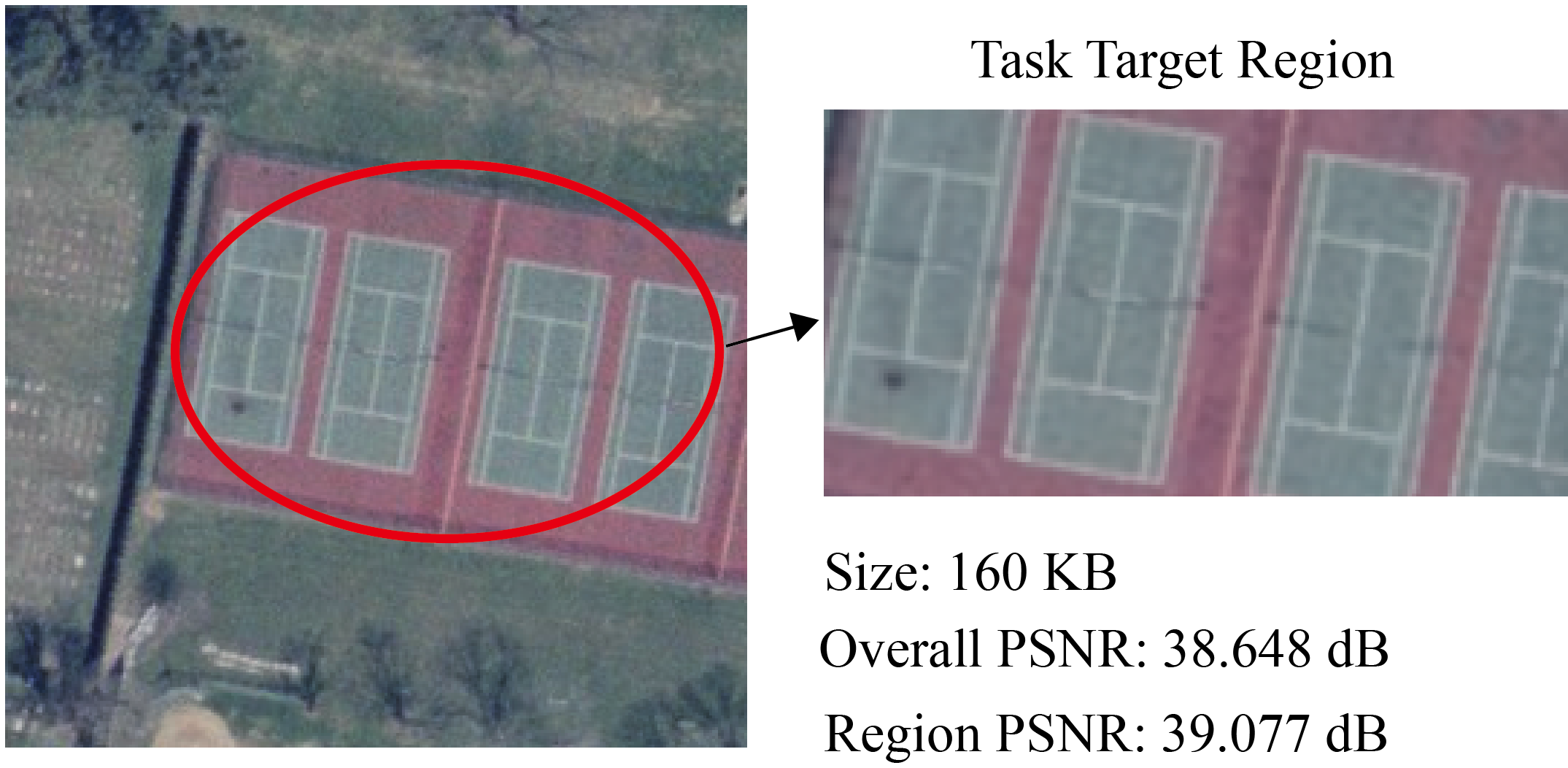}}
  \subfigure[Fuzzy filtering is employed.]{
  \label{fig5b}
  \includegraphics[width=0.5\textwidth]{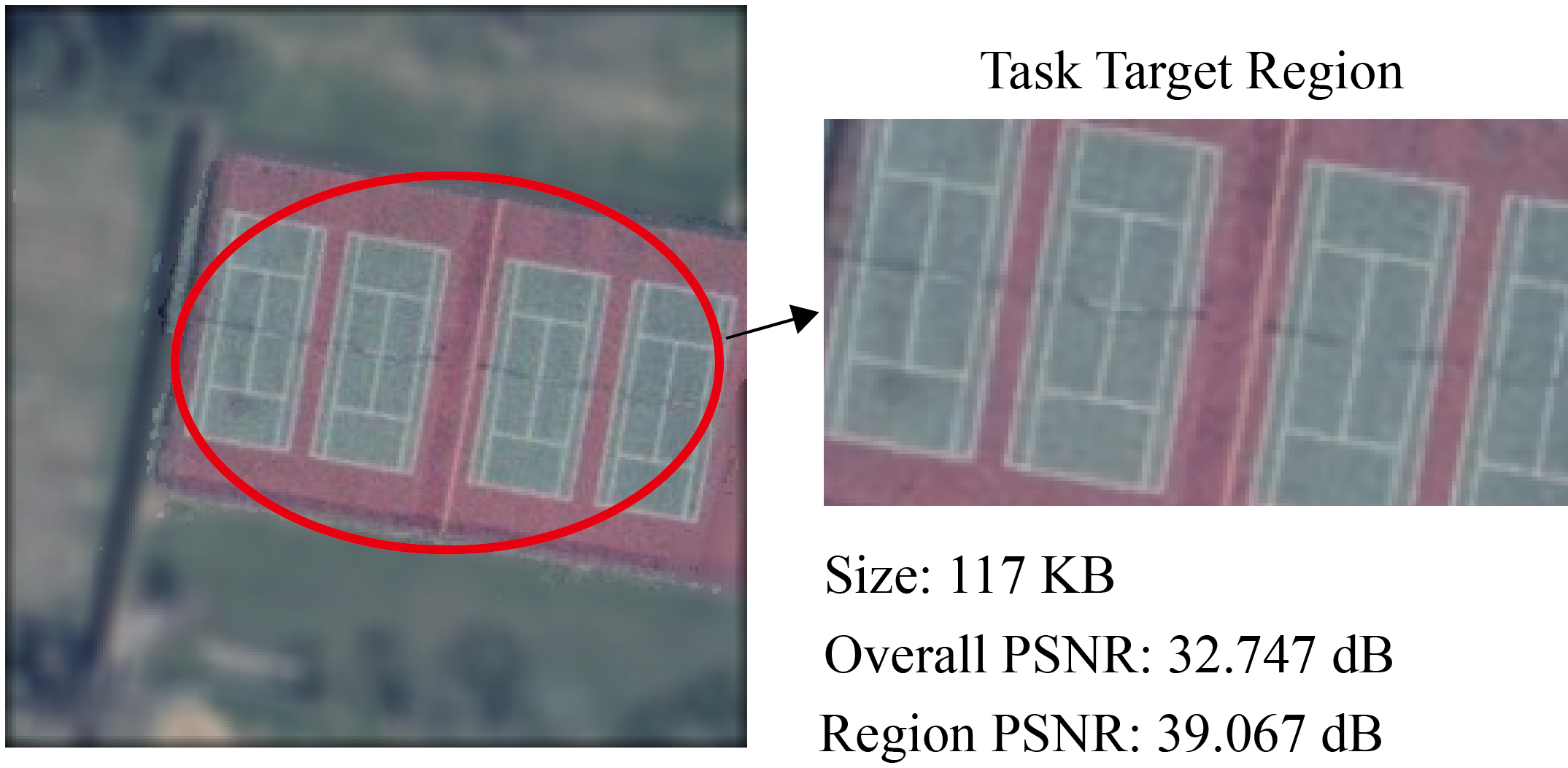}}
  \caption{Comparison of the effects of using fuzzy filtering or not. After using fuzzy filtering, the amount of transmitted data is greatly reduced while ensuring that the transmission quality of the task area is not reduced.}
  \label{fig5}
\end{figure}

Figure \ref{fig6} illustrates a performance comparison of image quality evaluation metrics under varying SNR conditions, focusing on the impact of adopting a stacking scheme. The control group employs a unified three-layer architecture trained throughout the SNR range without segmentation. In contrast, the experimental group utilizes a stacking strategy specifically optimized for the low SNR regime. This comparative analysis enables an assessment of the effectiveness of targeted architectural adaptation to enhance image quality under constrained transmission conditions.

Figure \ref{fig6a} presents the trend of variation of PSNR between different levels of SNR. In medium and high SNR regions, where noise interference is minimal, both schemes demonstrate comparable denoising performance, with closely aligned PSNR values. However, in low SNR conditions, the stacking scheme shows a clear advantage, improving PSNR by approximately 1.2 to 1.5 dB, which substantiates its robustness in high-noise scenarios. Figure \ref{fig6b} illustrates the SSIM index as a function of SNR. The results indicate that under low SNR conditions, the stacked IRST scheme consistently preserves structural similarity better than the baseline approach, evidenced by a significantly higher SSIM score. This suggests enhanced reconstruction of image details and structural information.

These results underscore the efficacy of the stacking architecture in low SNR environments. The design integrates a multilayer semantic reconstruction module that enables adaptive learning across varying noise intensities, thereby strengthening the model's capability to cope with complex channel conditions while preserving architectural cohesion. In contrast, the control group adopts a uniform training structure, which performs competitively in high SNR conditions. However, it lacks specialized optimization for low SNR scenarios, so its performance noticeably degrades in high-noise environments.

\begin{figure}[tbp]
  \centering
  \subfigure[The comparison of PSNR.]{
  \label{fig6a}
  \includegraphics[width=0.5\textwidth]{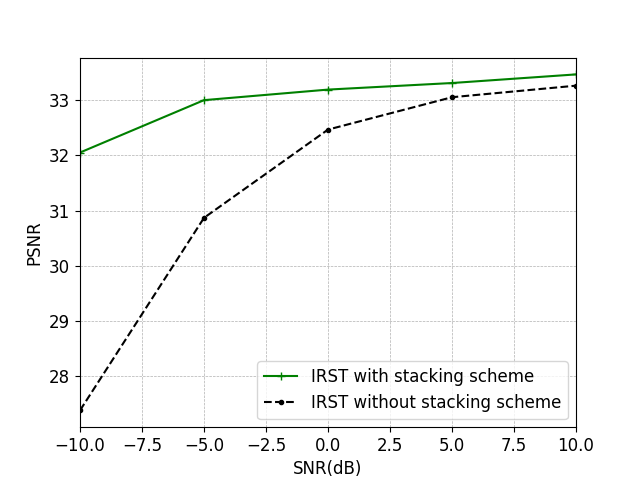}}
  \subfigure[The comparison of SSIM.]{
  \label{fig6b}
  \includegraphics[width=0.5\textwidth]{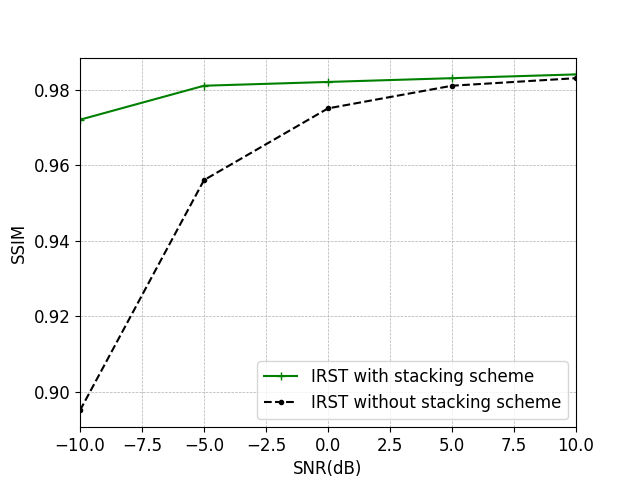}}
  \caption{PSNR and SSIM comparison under varying SNR conditions between IRST with stacking scheme and baseline without stacking optimization.}
  \label{fig6}
\end{figure}

\subsubsection{Performance Comparison}
Figure \ref{fig7} presents a comparative evaluation of image quality metrics across various SNR conditions for multiple image transmission schemes. Specifically, Figure \ref{fig7a} illustrates the PSNR performance trends. PSNR values increase with rising SNR levels for all methods, with the proposed IRST scheme demonstrating consistent superiority across the entire SNR spectrum. Notably, it achieves nearly 30 dB under low SNR conditions, underscoring its resilience in noisy environments. 

Figure \ref{fig7b} depicts SSIM variation across SNR levels. The IRST scheme, employing a stacked architecture, maintains high SSIM values from low to high SNR, indicating enhanced structural fidelity. Moreover, its SSIM values exhibit minimal fluctuation throughout the SNR range, reflecting robust adaptability. This stability arises from the dynamic participation of constituent models: under low SNR, more models are engaged during channel encoding and decoding to improve robustness; under high SNR, fewer models are activated, thereby reducing computational complexity and latency. 

Figure \ref{fig7c} further complements the evaluation by presenting LPIPS trends across varying SNR levels. The proposed IRST scheme consistently achieves the lowest LPIPS scores across the entire SNR range, highlighting its superior perceptual fidelity. Even under severely degraded conditions (e.g., SNR = - 10 dB), IRST maintains a perceptual advantage over competing methods such as SS-DDPM, Deep JSCC, and WITT. This performance underscores IRST’s ability to preserve visually meaningful features despite channel noise, further validating the effectiveness of its adaptive model stacking strategy.

\begin{figure}[tbp]
  \centering
  \subfigure[PSNR performance comparison of different schemes.]{
  \label{fig7a}
  \includegraphics[width=0.5\textwidth]{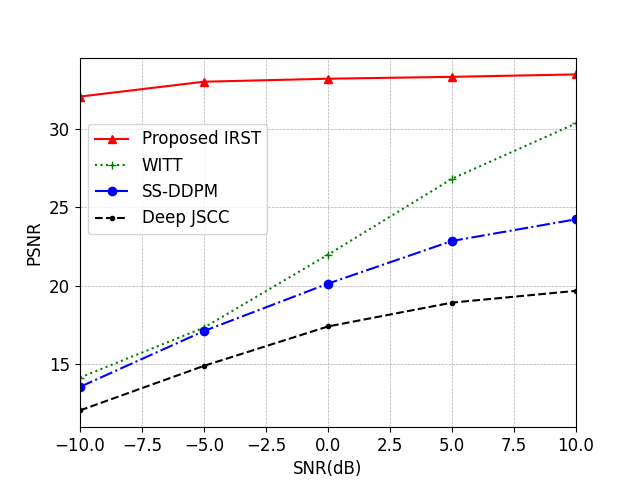}}
  \subfigure[SSIM performance comparison of different schemes.]{
  \label{fig7b}
  \includegraphics[width=0.5\textwidth]{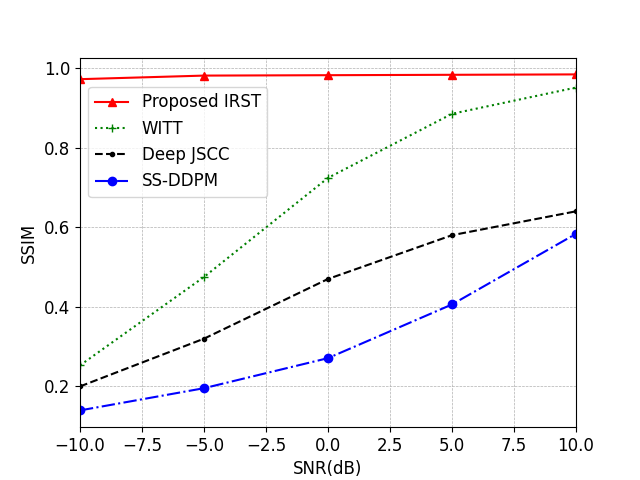}}
  \subfigure[LPIPS performance comparison of different schemes.]{
  \label{fig7c}
  \includegraphics[width=0.5\textwidth]{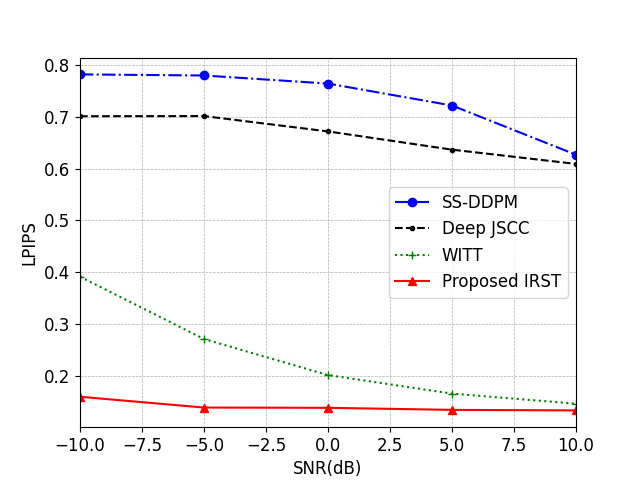}}
  \caption{The performance comparison of different schemes.}
  \label{fig7}
\end{figure}

\subsubsection{Adaptation Comparison}
Figure \ref{fig8} presents experimental results on heterogeneous datasets. The model is trained using the DLRSD remote sensing image dataset. For evaluation, the NWPU VHR-10 dataset is used, which differs markedly in structure and resolution. This setup enables assessment of the model’s adaptability and robustness in cross-domain scenarios.

Figure \ref{fig8a} illustrates the PSNR outcomes of four image transmission schemes under varying SNR conditions. Across the entire SNR spectrum, the proposed IRST method consistently yields the highest PSNR, demonstrating strong resilience to noise and high fidelity in image reconstruction. The WITT method ranks second, particularly excelling in medium and high SNR regions. SS-DDPM, though slightly less performant, still exhibits noticeable denoising capability. In contrast, Deep JSCC is limited by its simplistic structure and inadequate semantic feature extraction, which suffers pronounced performance degradation in low SNR conditions.

Figure \ref{fig8b} displays the SSIM curves under the same SNR settings. The IRST scheme consistently achieves superior SSIM across most SNR intervals, indicating its strong capacity to preserve perceptual quality despite dataset heterogeneity. WITT and SS-DDPM maintain substantial structural consistency in high SNR ranges, while Deep JSCC exhibits insufficient structural retention under low SNR, reflecting limited robustness.

Figure \ref{fig8c} presents the LPIPS performance of the four image transmission schemes under varying SNR conditions. The proposed IRST method consistently achieves the lowest LPIPS values throughout the SNR range, indicating superior perceptual similarity and robustness to noise. WITT is second, particularly in the medium to high SNR regions, where it maintains competitive perceptual quality. SS-DDPM demonstrates moderate performance, showing effective denoising but limited perceptual fidelity in low SNR scenarios. Meanwhile, Deep JSCC also exhibits the high LPIPS scores, especially under low SNR conditions, reflecting its limited ability to preserve semantic and perceptual features due to its simplistic architecture.

Overall, the experimental findings affirm that the IRST architecture possesses robust generalization across datasets. It effectively sustains image quality under diverse channel conditions and data discrepancies, highlighting its suitability for real-world deployment in remote sensing communication. In contrast, the limited capacity of Deep JSCC to extract heterogeneous semantic features contributes to its poor adaptability when tested outside of its training domain.

\begin{figure}[tbp] 
  \centering
  \subfigure[PSNR performance comparison in using different dataset.]{
  \label{fig8a}
  \includegraphics[width=0.5\textwidth]{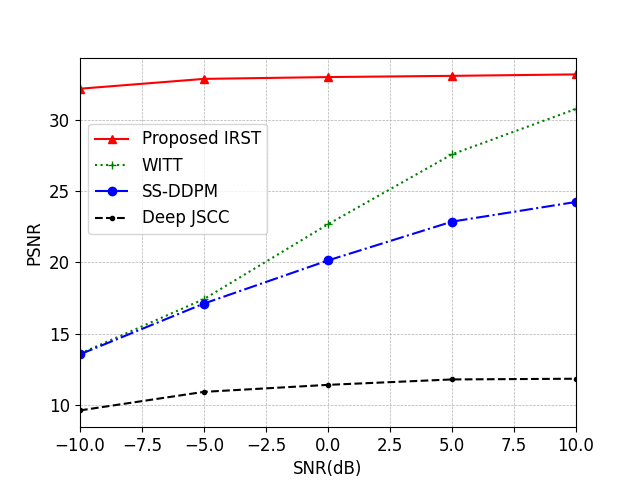}}
  \subfigure[SSIM performance comparison in using different dataset.]{
  \label{fig8b}
  \includegraphics[width=0.5\textwidth]{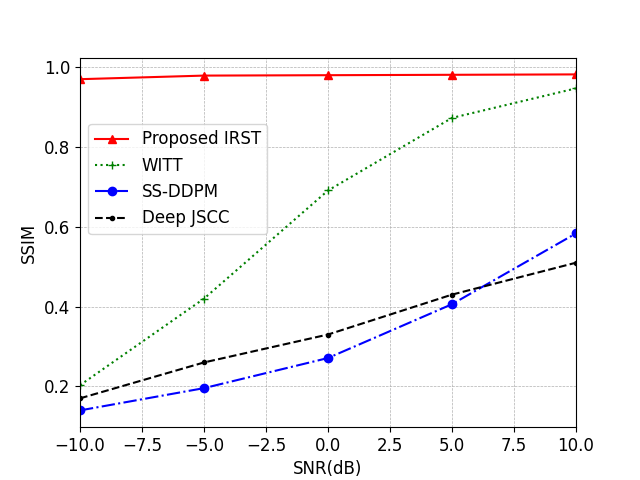}}
  \subfigure[LPIPS performance comparison in using different dataset.]{
  \label{fig8c}
  \includegraphics[width=0.5\textwidth]{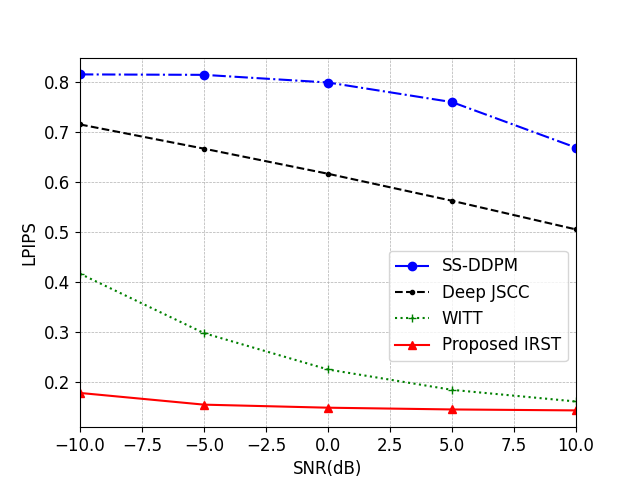}}
  \caption{The adaptation comparison of different schemes while train dataset is DLRSD and test dataset is NWPU VHR-10.}
  \label{fig8}
\end{figure}

\subsubsection{Visualization Result}
To enable a more comprehensive evaluation of transmission and reconstruction performance across different schemes, a visual comparison is presented in Figure \ref{fig9}.

\begin{figure*}[!t]
\centerline{\includegraphics[width=1\textwidth]{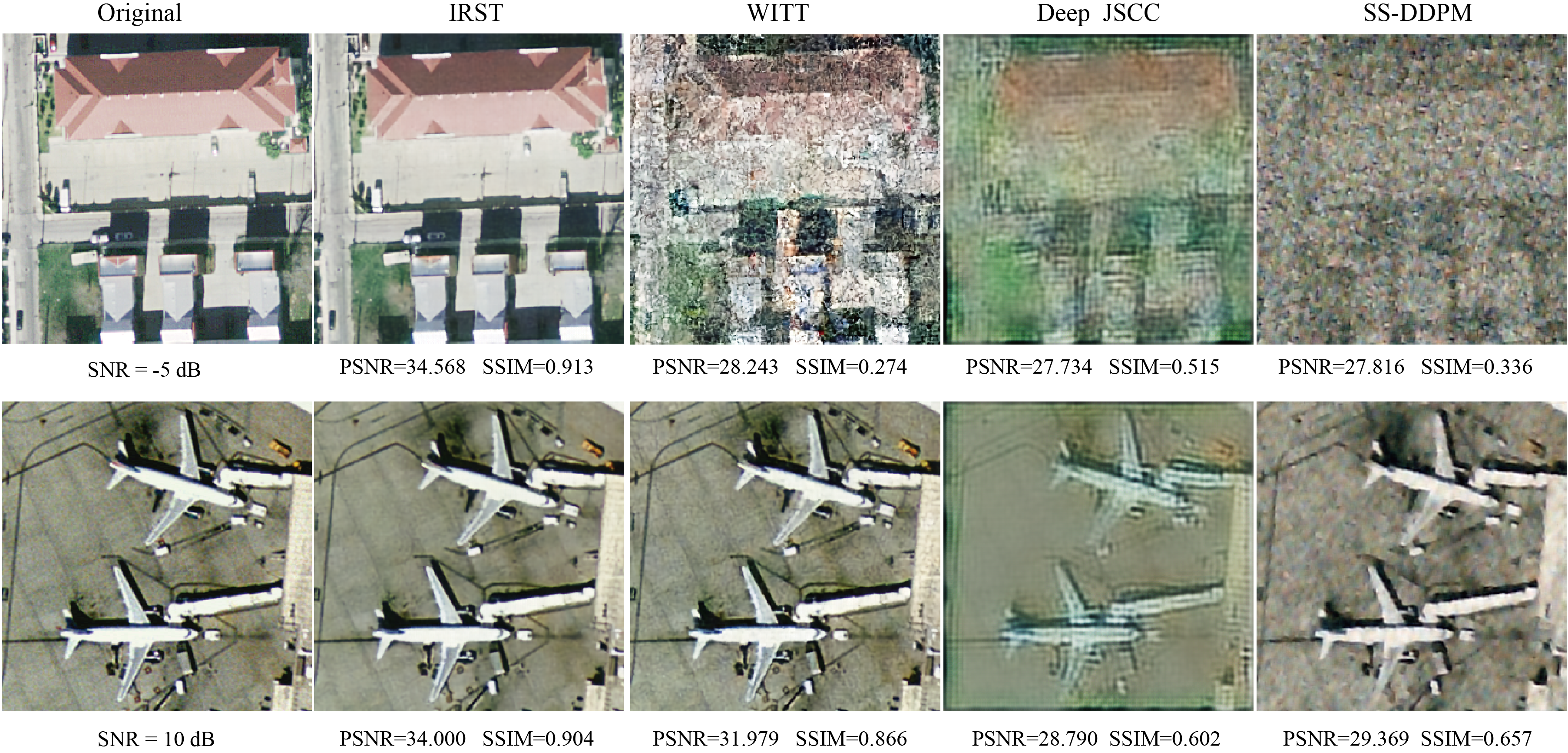}}
\caption{The visual comparison of different schemes. The first row depicts residential areas under low SNR environments, whereas the second row showcases an airport scene with high SNR. Across both scenarios, the proposed IRST model exhibits strong image restoration performance, effectively adapting to diverse noise intensities and structural complexities. This adaptability underscores its robustness and suitability for dynamic transmission conditions.}
\label{fig9}
\end{figure*}

The first row illustrates reconstruction outcomes for a residential scene under a low SNR condition of -5 dB. In this noise-intensive setting, most methods struggle to recover accurate structural information. The SS-DDPM method exhibits the weakest performance, producing only coarse outlines with considerable loss of detail, underscoring its limited capability in capturing semantic structures under high-noise conditions. Deep JSCC yields heavily distorted results, marked by pronounced color shifts in local regions. Although the WITT method manages to reconstruct the general scene structure, the finer details remain indistinct. In contrast, the proposed IRST method demonstrates superior reconstruction quality, successfully restoring building contours and textural details. This suggests enhanced noise resistance and semantic preservation.

The second row corresponds to an airport scene observed under a higher SNR condition of 10 dB. Under this more favorable environment, all methods exhibit improved performance, with SSIM values exceeding 0.7, indicating good structural consistency. Nonetheless, Deep JSCC continues to suffer from structural inaccuracies and color distortions, rendering it ineffective in restoring large-scale building features. The IRST method maintains robust performance, delivering complete structural reconstruction alongside clear visual details and well-defined edges. Although SS-DDPM and WITT perform comparably well in this high SNR setting, their results remain slightly inferior to those produced by IRST.

Overall, Figure \ref{fig9} convincingly demonstrates that the IRST method offers superior adaptability and consistent reconstruction quality across varying SNR levels and scene complexities. Its ability to sustain high image fidelity under low SNR conditions highlights the architectural advantages of IRST, particularly in terms of interference resilience and cross-scene semantic generalization.

\subsubsection{Complexity Comparison}

Table \ref{Table.5} presents a comparative analysis of model complexity across different architectures. The first row reports the memory footprint associated with model parameters, while the second row details the number of Multiply-Accumulate Operations (MACs), and the third line reports the GPU running time. As illustrated, both Deep JSCC and SS-DDPM primarily rely on CNN, leveraging parameter sharing and local receptive fields. This design results in relatively low memory consumption and computational cost. However, such architectures tend to struggle with preserving resolution when processing high-definition images under low SNR conditions. Furthermore, since SS-DDPM is based on a diffusion model for image reconstruction, its runtime is significantly longer than other methods.

In contrast, IRST and WITT utilize the Swin Transformer architecture, which incorporates hierarchical feature extraction and cross-regional modeling. These mechanisms significantly increase both memory usage and computational complexity, yet yield superior generative performance. Notably, IRST employs a dynamic SNR-aware channel codec, enabling more efficient memory utilization compared to WITT, which uses separately designed codecs for different channel conditions. Furthermore, because IRST skips redundant coding paths under high SNR conditions, it significantly reduces runtime, making it suitable for resource-constrained scenarios.

Overall, the proposed IRST model achieves a balanced trade-off between complexity and output quality, maintaining computational stability while adapting effectively to varying channel environments.

\begin{table}[]
\caption{Complexity Comparison}
\begin{tabular}{ccccc}

\hline
Model              & IRST  & WITT  & Deep JSCC & SS-DDPM \\ \hline
Parameter Size (M) & 19.89 & 21.72 & 1.56     & 5.11   \\
FLOPs (GMac)       & 32.45 & 32.82 & 2.62     & 6.73   \\
GPU Runtime (s)    & 4.97 & 6.79 & 5.51     & 8.04   \\      
\hline
\end{tabular}
\label{Table.5}
\end{table}

\section{Conclusion}
\label{section7}
The integration of SemCom with satellite networks offers significant potential for future intelligent communication systems. In this paper, we propose the IRST model in the scenario of limited bandwidth and dynamic channel changes. In this model, an SME scheme is first proposed to improve the accuracy of semantic segmentation. Then, based on the task requirements and the current channel conditions, we propose a semantic selection scheme to select the transmission content. Meanwhile, we design a stack-based SNR channel codec, which can perform dynamic channel coding according to the current SNR. Finally, we compare the proposed model with other schemes and the results show that the IRST model performs well under various conditions.

Although the IRST model proposed in this paper demonstrates significant improvements in the robustness and efficiency of satellite-ground SemCom, several limitations remain that warrant further investigation. First, the current validation is performed primarily on remote sensing image tasks, and the model adaptability to other modalities (such as speech, text, and video) has not been thoroughly explored, thus constraining its generalization capability. Moreover, channel modeling and feedback mechanisms are based on idealized assumptions, without fully accounting for the dynamic variations and latency inherent in real-world satellite-ground links, including Doppler effects and energy consumption, which may compromise the accuracy of semantic selection. Future research directions include extending the IRST framework to support multimodal data in order to enhance system versatility; incorporating reinforcement learning and adaptive strategies to enable end-to-end semantic optimization and dynamic policy adjustment; and conducting deployment verification in real world or high-fidelity simulated satellite-ground environments to rigorously assess the stability and practical applicability of the model under complex conditions.

\bibliography{arxiv} 

\bibliographystyle{IEEEtran}

% \newpage

 

% \end{IEEEbiography}



\vfill

\end{document}